\documentclass[a4paper,11pt]{article}
\pdfoutput=1 % if your are submitting a pdflatex (i.e. if you have
\usepackage{jheppub} % for details on the use of the package, please
\usepackage{amsmath, amsfonts}

\usepackage[T1]{fontenc} % if needed
\usepackage{xcolor}
\usepackage{subfigure}
\usepackage{stmaryrd}
\usepackage{enumerate}
\usepackage{tikz}
\usetikzlibrary{calc,intersections,positioning,arrows}
\usepackage[english]{babel}
\usepackage{graphicx}
\usepackage{bm}
\usepackage{epsf}

\renewcommand{\vec}[1]{\bm{#1}}

\newcommand{\beq}{\begin{equation}}
\newcommand{\eeq}{\end{equation}}
\newcommand{\bea}{\begin{eqnarray}}
\newcommand{\eea}{\end{eqnarray}}

\DeclareMathOperator{\Tr}{Tr}

\usepackage{color}

\usepackage[normalem]{ulem}  % \sout{old text} for strikeout

\newcommand{\non}{\nonumber\\}

\newcommand{\be}{\begin{equation}}
\newcommand{\ee}{\end{equation}}

\title{\boldmath Chiral anomaly induces superconducting baryon crystal}
\author{Geraint W.\ Evans}
\author{and Andreas Schmitt}

\affiliation{Mathematical Sciences and STAG Research Centre, University of Southampton, Highfield Campus, Southampton
SO17 1BJ, United Kingdom.}

\emailAdd{g.w.evans@soton.ac.uk}
\emailAdd{a.schmitt@soton.ac.uk}

\abstract{It was previously shown within chiral perturbation theory that the ground state of QCD in a sufficiently large magnetic field and at nonvanishing, but not too large, baryon chemical potential is a so-called chiral soliton lattice. The crucial ingredient of this observation was the chiral anomaly in the form of a Wess-Zumino-Witten term, which couples the baryon chemical potential to the magnetic field and the gradient of the neutral pion field. It was also shown that the chiral soliton lattice becomes unstable towards charged pion condensation at larger magnetic fields. 
We point out that this instability bears a striking resemblance to the second critical magnetic field of a type-II superconductor, however with the superconducting phase appearing upon {\it increasing} the magnetic field. The resulting phase has a periodically varying charged pion condensate that coexists with a neutral pion supercurrent. We construct this phase analytically in the chiral limit and show that it is energetically preferred. Just like an ordinary type-II superconductor, it exhibits a hexagonal array of magnetic flux tubes, and, due to the chiral anomaly, a spatially oscillating baryon number of the same crystalline structure. 
}

\begin{document} 
\maketitle
\flushbottom

%%%%%%%%%%%%%%%%%%%%%%%%%%%%%%%%%%%%%%%%%%%%%%%%%%%%%%%%%%%%%%%%%%%%%
\section{Introduction}
\label{sec:intro}
%%%%%%%%%%%%%%%%%%%%%%%%%%%%%%%%%%%%%%%%%%%%%%%%%%%%%%%%%%%%%%%%%%%%%

\begin{figure}
    \centering
    \includegraphics[width=0.65\textwidth]{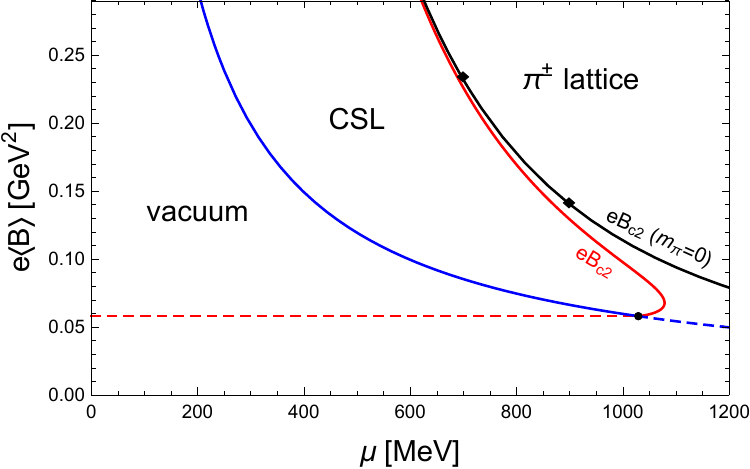}
    \caption{Phase diagram in the $e\langle B\rangle$-$\mu$ plane for a physical pion mass (red and blue curves) and in the chiral limit (black curve; in this case the vacuum only exists on the two axes). The solid blue curve marks the transition between the vacuum and the CSL phase. Just above the curve it is favourable to place a single domain wall into the system. The solid red curve indicates the instability of the CSL phase towards charged pion condensation at the critical field $B_{c2}$. In this paper,  we construct the resulting charged pion lattice in the chiral limit, in which case the critical field is given by the black curve. The two diamonds indicate the points at which 
we shall later 
    plot the lattice structures. The dashed curves (instability of a single domain wall, red, and continuation of the vacuum/CSL transition curve, blue) are transitions between metastable phases.  A first-order transition between the vacuum and the $\pi^\pm$ lattice is
    expected somewhere between the dashed lines. Actual baryons in the form of nuclear matter are expected to compete with and possibly replace the phases at sufficiently large $\mu$, but are omitted here and in the rest of the paper for simplicity. To put the scale of the magnetic field into context, note that within natural Heaviside-Lorentz units $e\langle B\rangle = 0.1\, {\rm GeV}^2\simeq 5.18 \, m_\pi^2 $ corresponds to $e\langle B\rangle \simeq 5.12\times 10^{18}\, {\rm G}$. }
    \label{fig:eBmu}
\end{figure}

Constructing the phase diagram of Quantum Chromodynamics (QCD) at nonvanishing baryon chemical potential $\mu$ is a very difficult problem. At sufficiently low temperatures, nuclear matter appears when the chemical potential becomes of the order of the nucleon mass, $\mu\sim 1\ {\rm GeV}$, and a transition to deconfined quark matter is expected at a larger -- unknown -- value of $\mu$. Both in nuclear and in quark matter additional phase transitions are expected due to various superfluid and superconducting phases \cite{Alford:2007xm}. In the presence of a magnetic field $B$, baryon number can appear for values of $\mu$ smaller than the nucleon mass -- albeit not in the form of ordinary nucleons. This is due to the chiral anomaly, the non-conservation of the axial current solely from quantum effects. The chiral anomaly gives rise to a coupling of the chemical potential to the magnetic field and the gradient of the neutral pion field, which changes the thermodynamics of the system qualitatively and induces a nonzero baryon number if all three quantities are nonzero \cite{Son:2004tq,Son:2007ny}. Such a phase with topological baryon number is indeed stable 
above a critical magnetic field $B={\rm const}/\mu$. It can be thought of as a stack of domain walls perpendicular to the magnetic field, such that baryon number oscillates in the direction of the magnetic field. This phase was termed Chiral Soliton Lattice (CSL) \cite{Brauner:2016pko}. In the chiral limit, i.e.\ neglecting the pion mass $m_\pi$, this phase exists for arbitrarily small nonzero $B$ and $\mu$ and has a uniform baryon density. In this limit, it was also discussed  in the framework of holography \cite{Thompson:2008qw,Rebhan:2008ur}. Within this framework, its interplay with nuclear matter was explored \cite{Preis:2011sp}, which was also investigated within a Skyrme model \cite{Kawaguchi:2018fpi,Chen:2021vou}. In analogy to an ordinary superfluid, where the gradient of a scalar field is related to the superfluid velocity, the pion gradient has also been referred to as a supercurrent \cite{Rebhan:2008ur}, similar to the kaon supercurrent phase in dense quark matter \cite{Kryjevski:2005qq,Schafer:2005ym}. We show the phase structure including the CSL phase in Fig.\ \ref{fig:eBmu}.

The CSL phase (as well as a single pion domain wall) becomes unstable towards charged pion condensation at a certain critical field, which was demonstrated with the help of pionic fluctuations  \cite{Son:2007ny,Brauner:2016pko}. So far, the phase beyond the instability, labelled "$\pi^\pm$ lattice" in Fig.\ \ref{fig:eBmu}, has not been constructed explicitly. It is the main goal of this paper to construct this phase and show that it is energetically favoured over the CSL phase above the critical field predicted in Ref.\ \cite{Brauner:2016pko}. In the chiral limit, this critical field behaves like $B= {\rm const}/\mu^2$, see black curve in Fig.\ \ref{fig:eBmu}. Since the charged pion condensate will turn out to be spatially inhomogeneous, the relevant thermodynamical variable used in Fig.\ \ref{fig:eBmu} and later in our calculation is the spatially averaged magnetic field $\langle B\rangle$ (multiplied by the elementary charge $e$). In the explicit calculation of this crystalline phase we shall restrict ourselves to the chiral limit, where the nontrivial lattice structure is two-dimensional, while 
in the presence of a nonzero pion mass a three-dimensional lattice is expected. In particular, the periodic structure of the magnetic field will, via the chiral anomaly, give rise to a crystalline structure for baryon number, such that the phase we construct is a superconducting baryon crystal.

One of our key observations is the similarity of the CSL instability with the instability of an ordinary electronic type-II superconductor at the second critical magnetic field \cite{Abrikosov:1957classic,Kleiner:1963Bulk,Essman:1967ExpTri,tinkham2004introduction}, usually referred to as $B_{c2}$, a notation we have adopted in Fig.\ \ref{fig:eBmu}. In the conventional scenario, the normal-conducting phase is preferred for $B>B_{c2}$, and Cooper pair condensation sets in just below $B_{c2}$. Type-II superconductivity allows for a partial expulsion of the magnetic field, in contrast to type-I superconductors, where the magnetic field is completely expelled due to the Meissner effect. This partial expulsion manifests itself in the formation of a lattice of magnetic flux tubes, together with a spatially varying Cooper pair condensate. Pioneered by Abrikosov \cite{Abrikosov:1957classic}, linearised Ginzburg-Landau theory can be employed to compute this lattice analytically just below $B_{c2}$ and to determine the preferred lattice structure. Our calculation is an application of these methods to chiral perturbation theory at the critical field of the CSL instability. The main differences to the textbook scenario are the presence of the neutral pions, whose interaction with the charged field is dictated by chiral perturbation theory, and the anomalous contribution. It turns out that superconductivity is induced by {\it increasing} the magnetic field, i.e.\ the normal-conducting CSL phase below $B_{c2}$ is superseded by the superconducting pion lattice above $B_{c2}$, inverting the order of phases in the non-anomalous, single-component superconductor.   

As in the related works \cite{Son:2007ny,Brauner:2016pko} we shall work within two-flavor chiral perturbation theory, where the chiral anomaly is implemented through a Wess-Zumino-Witten (WZW) term \cite{WessZumino:1971WZW,Witten:1983WZW}, and only consider mesonic degrees of freedom. This should be kept in mind for the interpretation of our results because, firstly, chiral perturbation theory is an effective theory of QCD at low energies, and we should stop trusting it literally for energies of the order of or larger than the typical scale for chiral symmetry breaking of about $4\pi f_\pi \sim 1\, {\rm GeV}$, where $f_\pi$ is the pion decay constant. 
As Fig.\ \ref{fig:eBmu} shows, our pion lattice exists in a regime close to or even above this scale. Secondly, actual baryons will play a role for large chemical potentials. At zero magnetic field, their onset is at $\mu\simeq 923\, {\rm MeV}$, and it is unknown from first principles how this onset changes with the magnetic field (for model calculations see for instance Refs.\ \cite{Preis:2011sp,Haber:2014ula}). For both reasons we have to treat our results with some care. They can be viewed as a prediction for a qualitatively novel phase, whose existence in QCD needs to be checked in the future with more elaborate methods. We should also emphasise that, while we present most of our derivations for a general pion mass, our main results are given for $m_\pi=0$ for simplicity. This is a good approximation for extremely large magnetic field, as the convergence of the red (physical) and black (chiral limit) curves in Fig.\ \ref{fig:eBmu} suggests. However, ultimately we are interested in the scenario with a physical pion mass and our study can be considered a first step towards this goal.

Our main motivation is of theoretical nature, having in mind a better understanding of the QCD phase diagram. One may ask if there is also a phenomenological motivation for our work. The magnitude of the magnetic fields discussed here is very large, perhaps too large to be relevant for any observational consequences. However, large magnetic fields are present in 
heavy-ion collisions and in the interior of neutron stars, possibly up to $B\sim (10^{18}-10^{20})\,\rm{G}$ \cite{Lai1991ColdEO,Kharzeev:2007jp,Ferrer:2010wz,Potekhin:2011eb,Kharzeev:2012ph,Adamczyk:2015eqo}. In the context of heavy-ion collisions our results are not directly applicable since we work at zero temperature, and finite-temperature extensions such as for the CSL phase in Ref.\ \cite{BraunerKolesovaYamamoto:2021WarmCSL} would be necessary. In the context of neutron stars, crystalline structures are of great interest, especially for potential observations of continuous gravitational waves due to "mountains" which can be sustained by a rigid structure in the interior of the star \cite{Glampedakis:2017nqy,LIGOScientific:2022lsr}. Therefore, our results might be of potential relevance for astrophysical observations. Of course, any conventional picture of a neutron star contains nuclear matter, and our results would have to be supplemented by the inclusion of baryonic degrees of freedom. Alternatively, one might speculate whether exotic stars with relatively small baryon chemical potential but large magnetic field may exist, where baryon number is purely generated by the chiral anomaly. As we shall see, the baryon numbers reached in the inhomogeneous phase constructed here are comparable to the ones expected inside neutron stars, i.e.\ of the order of and larger than nuclear saturation density. 

Finally, let us put our study in the context of other inhomogeneous superconductors proposed as candidate phases for the QCD phase diagram. Perhaps most closely related to our work is the charged pion condensate at nonzero isospin chemical potential, which has been 
studied within lattice QCD in its homogeneous version without a magnetic field \cite{Brandt:2017oyy}. If a magnetic field is switched on, a flux tube array is expected since the pion condensate turns out to be a type-II superconductor \cite{Adhikari:2015wva}. This inhomogeneous phase has been constructed within the same framework as used here \cite{Adhikari:2018Solo,Adhikari:2018Choi}, however its appearance does not rely on the chiral anomaly, neither does it show the inverted behaviour at $B_{c2}$. An analogue of the CSL phase at nonzero isospin chemical potential due to the axial anomaly does exist as well and its competition with the charged pion condensation has been studied recently \cite{Gronli:2022cri}, see also Ref.\ \cite{Adhikari:2015pva}. An inhomogeneous pion condensate with electromagnetic supercurrents has been constructed in  
Ref.\ \cite{Canfora:2020kyj,Canfora:2020uwf,Barriga:2021eki}, however without external magnetic field and without identifying its possible relevance for the QCD phase diagram. The scenario where a superconducting phase occurs above rather than below a critical field was proposed for 
charged rho meson condensation, where an Abrikosov lattice was also predicted \cite{Chernodub:2010RhoCond, Chernodub:2011}. 
Magnetic flux tube lattices similar to the one considered here  are also expected  in the QCD phase diagram at larger baryon chemical potentials, in nuclear matter due to Cooper pairing of protons \cite{Alford:2007np,Haber:2017kth,Wood:2020czv} and in colour-superconducting quark matter, where the multi-component structure of the system gives rise to unconventional flux tubes \cite{Haber:2017oqb,Haber:2018tqw,Evans:2020uui}. 

Our paper is structured as follows. We start by establishing our 
formalism within chiral perturbation theory in Sec.\ \ref{sec:setup}. 
After the formulation of the Lagrangian in Sec.\ \ref{sec:lag}, this 
includes the derivation of the general form of the equations of motion and the free energy in Sec.\ \ref{sec:eom}, and a brief recapitulation of the instability of the CSL phase in Sec.\ \ref{sec:CSL}. Our main results are derived in Sec.\ \ref{sec:EoMs}, where, firstly, we 
present the expansion at the critical field for the general case in Sec.\ \ref{sec:exp0}, and then derive and discuss the flux tube lattice in the chiral limit in Sec.\ \ref{sec:chiral}. For both main sections, in particular Sec.\ \ref{sec:EoMs}, it is useful to be familiar with the standard treatment of a type-II superconductor within a Ginzburg-Landau approach, which we recapitulate in Appendix \ref{appB} and which the reader may consult as a warm-up for the main part. We give a summary and an outlook in Sec.\ \ref{sec:out}. 
Throughout the paper, our convention for the Minkowski metric is  $g^{\mu\nu} = \text{diag}\,(1,-1,-1,-1)$ and we work in natural units where $\hbar=c=k_{B}=1$. For the electromagnetic part we use Heaviside-Lorentz units, such that the elementary charge is  $e=\sqrt{4\pi\alpha}\simeq0.3028$ with the fine structure constant $\alpha$. 

%%%%%%%%%%%%%%%%%%%%%%%%%%%%%%%%%%%%%%%%%%%%%%%%%%%%%%%%%%%%%%%%%%%%%
\section{Setup and CSL instability}
\label{sec:setup}
%%%%%%%%%%%%%%%%%%%%%%%%%%%%%%%%%%%%%%%%%%%%%%%%%%%%%%%%%%%%%%%%%%%%%

%%%%%%%%%%%%%%%%%%%%%%%%%%%%%%%%%%%%%%%%%%%%%%%%%%%%%%%%%%%%%%%%%%%%%
\subsection{Lagrangian}
\label{sec:lag}
%%%%%%%%%%%%%%%%%%%%%%%%%%%%%%%%%%%%%%%%%%%%%%%%%%%%%%%%%%%%%%%%%%%%%

Our starting point is the Lagrangian containing an electromagnetic part, a chiral part, and a WZW part,
\bea \label{Lfull}
{\cal L} = {\cal L}_{\rm em} + {\cal L}_\Sigma+ \mathcal{L}_{\text{WZW}} \, .
\eea
The electromagnetic part is 
\be \label{Lem}
{\cal L}_{\rm em} = -\frac{1}{4}F_{\mu\nu}F^{\mu\nu} \, , 
\ee
where $F^{\mu\nu}=\partial^\mu A^\nu - \partial^\nu A^\mu$ is the electromagnetic field strength tensor with the electromagnetic gauge field $A^\mu$. The chiral part is the usual leading-order chiral Lagrangian \cite{GasserLeutwyler:1983CPT,Ecker:1994gg}
\begin{equation}
    \mathcal{L}_\Sigma = \frac{f_\pi^2}{4}\mathrm{Tr}\left[\nabla_\mu \Sigma^{\dagger}\nabla^{\mu}\Sigma\right] + \frac{m_\pi^2 f_\pi^2}{4}\mathrm{Tr}\left[ \Sigma + \Sigma^{\dagger} \right] \, ,
    \label{ChiralLagrangian}
\end{equation}
where 
\be \label{Sigma1}
\Sigma = e^{i\phi_a\tau_a}  = \frac{\sigma+i\pi_a\tau_a}{f_\pi} 
\ee
is the chiral $SU(2)$ field, with the Pauli matrices $\tau_a$, $a=1,2,3$,
and 
\be
\frac{\sigma}{f_\pi} = \cos\phi \, , \qquad \frac{\pi_a}{f_\pi} = \frac{\phi_a}{\phi}\sin\phi \, , 
\ee
where $\phi^2\equiv\phi_1^2+\phi_2^2+\phi_3^2$, such that $f_\pi^2 = \sigma^2+\pi_a\pi_a$. The covariant derivative is 
\bea \label{covariant}
\nabla^{\mu}\Sigma = \partial^{\mu}\Sigma -i\left[\mathcal{A}^{\mu},\Sigma \right] \, , 
\eea
with $\mathcal{A}^{\mu} = A^\mu_B+eQA^\mu$, where the auxiliary gauge field $A_B^\mu = (\mu,0,0,0)$ contains the baryon chemical potential, and $Q = {\rm diag}\,(2/3,-1/3)$ is the generator of the electromagnetic gauge group. In the covariant derivative, the gauge field contribution proportional to the unit matrix obviously drops out due to the commutator, such that ${\cal L}_\Sigma$ does not depend on $\mu$.  
Evaluating the traces and using the parametrisation (\ref{Sigma1}) the chiral Lagrangian can be written as 
\bea \label{LSig1}
{\cal L}_\Sigma &=&  \frac{1}{2}\partial_\mu\pi_0\partial^\mu\pi_0+D_\mu\varphi(D^\mu\varphi)^*+\frac{1}{2}\partial_\mu\sigma\partial^\mu\sigma
 +m_\pi^2f_\pi\sigma  \, , 
\eea
where the neutral pion field has been denoted by $\pi_0\equiv \pi_3$, where the charged pions have been combined in the complex scalar field
\be
\varphi = \frac{1}{\sqrt{2}}(\pi_1+i\pi_2) \, ,
\ee
and where we have defined the covariant derivative
$D^\mu\varphi = \partial^\mu\varphi+ieA^\mu\varphi$. 

For our purpose it will be convenient to reparametrise the chiral field, following Appendix A of Ref.\ \cite{Brauner:2016pko}. To this end, we separate the third component as follows,
\be \label{SigmaU}
\Sigma = e^{i\alpha\tau_3} U \, . 
\ee
Parametrising $U$ by the new fields $\sigma_0$, $\pi_1'$, $\pi_2'$,
\be \label{Usig}
U = \frac{\sigma_0+i(\pi_1'\tau_1+\pi_2'\tau_2)}{f_\pi} \, ,
\ee
and comparing both sides of Eq.\ (\ref{SigmaU}) component by component yields the explicit form of the transformation 
\bea
\begin{array}{rl}
\sigma = &\sigma_0\cos\alpha\\
\pi_0= &\sigma_0\sin\alpha
\end{array} \,\, , \qquad 
\begin{array}{rl}
\pi_1 = &\pi_1'\cos\alpha + \pi_2'\sin\alpha  \\
\pi_2 =& -\pi_1'\sin\alpha + \pi_2'\cos\alpha 
\end{array} \, \,,
\eea
i.e. it amounts to going from cartesian to polar coordinates in the $(\sigma,\pi_0)$ sector and applying a rotation by $\alpha$ in the $(\pi_1,\pi_2)$ sector. As a consequence, we can define a new complex field 
\be \label{phiprime}
\varphi' = \frac{1}{\sqrt{2}}(\pi_1'+i\pi_2') \, ,
\ee
which obeys the transformation $\varphi = e^{-i\alpha} \varphi'$. 
In the new coordinates we have the constraint $f_\pi^2 = \sigma_0^2+\pi_1'^2+\pi_2'^2 = \sigma_0^2+2|\varphi'|^2$, which ensures the unitarity of $U$. 
We can now write the Lagrangian (\ref{LSig1}) as 
\bea \label{Lagtrans}
{\cal L}_\Sigma &=& \frac{\sigma_0^2}{2}\partial_\mu\alpha\partial^\mu\alpha+D_\mu \varphi'(D^\mu\varphi')^*+\frac{1}{2}\partial_\mu\sigma_0\partial^\mu\sigma_0 +m_\pi^2f_\pi\sigma_0\cos\alpha \, ,
\eea
with the covariant derivative redefined as $D^\mu\varphi' = \partial^\mu\varphi'+i(eA^\mu-\partial^\mu\alpha)\varphi'$. We can eliminate the spurious field $\sigma_0$, such that our pionic degrees of freedom are the scalar field $\alpha$ (related to the neutral pion field in the absence of charged pions by $\pi_0 = f_\pi\sin\alpha$) and the complex field $\varphi'$ (related to the original charged pion fields by a rotation by the angle $\alpha$). The main benefit of the reparametrisation is that $\nabla\alpha$ will turn out to be constant in our crystalline phase in the chiral limit, which facilitates the calculation. From now on we shall drop the prime (we shall never come back to the original parametrisation), to arrive at the final form 
\bea  \label{LSig2}
{\cal L}_\Sigma &=&
D_\mu \varphi(D^\mu\varphi)^* + \frac{\partial_\mu|\varphi|^2\partial^\mu|\varphi|^2}{2(f_\pi^2-2|\varphi|^2)}  + \frac{f_\pi^2-2|\varphi|^2}{2}\partial_\mu\alpha\partial^\mu\alpha\non[2ex]
&& +m_\pi^2f_\pi\sqrt{f_\pi^2-2|\varphi|^2}\cos\alpha \, . 
\eea
One can check that this result is also obtained by inserting the chiral field (\ref{SigmaU}) directly into the Lagrangian (\ref{ChiralLagrangian}). 

The final term of the Lagrangian (\ref{Lfull}) accounts for the chiral anomaly through a WZW term \cite{WessZumino:1971WZW,Witten:1983WZW}, which we can write as
\begin{equation}
    \mathcal{L}_{\text{WZW}} = \left( A^B_{\mu} - \frac{e}{2}A_{\mu}\right)j^{\mu}_B\,,
    \label{ActionWZW}
\end{equation}
with the Goldstone-Wilczek baryon current \cite{Goldstone:1981kk,Son:2007ny,Brauner:2016pko}
\bea
    j^{\mu}_{B} &=& -\frac{\epsilon^{\mu\nu\rho\lambda}}{24\pi^2}\text{Tr}\left[(\Sigma\nabla_{\nu}\Sigma^{\dagger})( \Sigma\nabla_{\rho}\Sigma^{\dagger})( \Sigma\nabla_{\lambda}\Sigma^{\dagger}) +\frac{3ie}{4}F_{\nu\rho}\tau_3\left( \Sigma\nabla_{\lambda}\Sigma^{\dagger} +\nabla_{\lambda}\Sigma^{\dagger}\Sigma\right)\right]\non[2ex]
   &=&  -\frac{\epsilon^{\mu\nu\rho\lambda}}{4\pi^2}\partial_{\nu}\alpha
   \left(\frac{e}{2}F_{\rho\lambda}+ \frac{\partial_{\rho}j_{\lambda}}{ef_{\pi}^2}
        \right)\,,
    \label{BCurrent}
\eea
where   
\be
    j^{\mu} = ie\left(\varphi^*\partial^{\mu}\varphi - \varphi\partial^{\mu}\varphi^*\right) -2e\left(eA^{\mu}-\partial^{\mu}\alpha\right)|\varphi|^2
    \label{QCurrent}
\ee
is the non-anomalous contribution to the charged current.
Details on the evaluation of the traces leading to the second line of Eq.\ \eqref{BCurrent} can be found in Appendix \ref{appA}. Interestingly, we see that besides the electromagnetic term, there is a vorticity contribution. These two terms are reminiscent of the chiral magnetic and chiral vortical effects, which are generated by the chiral anomaly and which have been studied extensively mostly in the context of heavy-ion collisions, see for instance Ref.\ \cite{Landsteiner:2016led} for an introductory review. A similar vorticity term  gives rise to the CSL phase in a rotating system even without magnetic field \cite{Huang:2017pqe}, in which case also inhomogeneous pion-condensed phases have been predicted \cite{Eto:2021gyy}. Here we do not impose any rotation on the system, but we shall see that the dynamically created charged current does render the vorticity term nonzero, which has a direct impact on the baryon number.

From the explicit result for the baryon current \eqref{BCurrent} we  immediately conclude its conservation, 
\be \label{jB0}
\partial_\mu j^\mu_B = 0 \, ,
\ee
as it should be. In summary, our total Lagrangian is given by Eqs.\ (\ref{Lem}), (\ref{LSig2}), and (\ref{ActionWZW}) with the baryon current (\ref{BCurrent}). 

%%%%%%%%%%%%%%%%%%%%%%%%%%%%%%%%%%%%%%%%%%%%%%%%%%%%%%%%%%%%%%%%%%%%%
\subsection{Equations of motion and free energy}
\label{sec:eom}
%%%%%%%%%%%%%%%%%%%%%%%%%%%%%%%%%%%%%%%%%%%%%%%%%%%%%%%%%%%%%%%%%%%%%

The equations of motion for $\varphi^*$, $\alpha$, $A_{\mu}$ become, respectively, 
\begin{subequations}\allowdisplaybreaks
    \bea
        0&=& \left[D_{\mu}D^{\mu} +\partial_{\mu}\alpha\partial^{\mu}\alpha +\frac{\partial_{\mu}\partial^{\mu}|\varphi|^2}{f_{\pi}^2-2|\varphi|^2} +\frac{\partial_{\mu}|\varphi|^2\partial^{\mu}|\varphi|^2}{\left(f_{\pi}^2-2|\varphi|^2\right)^{2}}
        +\frac{m_{\pi}^2 f_{\pi}\cos{\alpha}}{\sqrt{f_{\pi}^2 -2|\varphi|^2}}\right.\non[2ex]
        &&
        \left.+\frac{ie\epsilon^{\mu\nu\rho\lambda}}{8\pi^2f_{\pi}^2}\partial_{\nu}\alpha\, F_{\rho\lambda}D_{\mu}\right]\varphi\,,
        \label{phiEOMFull}
        \\[2ex]
       0&=& \partial_\mu\left[(f_{\pi}^2 -2|\varphi|^2)\partial^{\mu}\alpha\right] +m_{\pi}^2f_{\pi}\sqrt{f_{\pi}^2 -2|\varphi|^2}\sin{\alpha} 
        -\frac{e\epsilon^{\mu\nu\rho\lambda}}{16\pi^2}F_{\mu\nu}\left( \frac{e}{2}F_{\rho\lambda}+\frac{\partial_{\rho}j_{\lambda}}{ef_{\pi}^2}\right)
        \label{alphaEOMFull}
        \,, \hspace{1cm}\\[2ex]        
       0 &=& -\partial_{\nu}F^{\nu\mu}+j^{\mu} +\frac{e}{2}j^{\mu}_B
        -\frac{e^2\epsilon^{\mu\nu\rho\lambda}}{16\pi^2}\partial_{\nu}\alpha \, F_{\rho\lambda}\left(1 -\frac{2|\varphi|^2}{f_{\pi}^2}\right)
        \label{gaugeEOMFull}
        \,.
    \eea
\end{subequations}
The last equation is an extended Maxwell equation, including anomalous contributions to the charged current. Making use of the 
baryon number conservation (\ref{jB0}) we find that the total electric charge conservation reads
\be
0 = \partial_\mu j^\mu+\frac{e^2\epsilon^{\mu\nu\rho\lambda}}{8\pi^2f_\pi^2}\partial_\mu|\varphi|^2\partial_\nu\alpha\,F_{\rho\lambda} \, .
\ee
This relation has been used in the derivation of the equation of motion for $\alpha$ (\ref{alphaEOMFull}).

We will only be interested in the static limit, and we assume the system to be locally charge neutral, which can be achieved for instance by adding a gas of electrons or positrons. This is very similar to the standard Ginzburg-Landau treatment of an electronic superconductor, where the negative charge of the electron Cooper pairs is cancelled by the surrounding lattice of ions. As a consequence, the electric field vanishes and we may ignore Gauss' law, i.e.\ the temporal component of Eq.\ (\ref{gaugeEOMFull}). For the remaining equations we can therefore ignore all time derivatives and set $A_0=0$. (Note that even for $\partial_t=A_0=0$ there is an anomalous electric charge contribution in Gauss' law which can only be ignored under the assumption of a neutralising lepton gas.) As a result, all anomalous contributions to the equations of motion vanish and we 
arrive at
\begin{subequations}
    \bea
        0&=&\left[{\cal D} +\frac{\Delta|\varphi|^2}{f_{\pi}^2-2|\varphi|^2} +\frac{\left(\nabla|\varphi|^2\right)^2}{\left(f_{\pi}^2-2|\varphi|^2\right)^{2}}+m_\pi^2\cos\alpha\left(1-\frac{f_\pi}{\sqrt{f_\pi^2-2|\varphi|^2}}\right) \right]\varphi\,, \hspace{0.5cm}
        \label{phiEOM}
        \\[2ex]
        0&=&\nabla\cdot\left[(f_\pi^2-2|\varphi|^2)\nabla\alpha\right] -m_{\pi}^2f_\pi\sqrt{f_\pi^2-2|\varphi|^2}\sin{\alpha}\,,
        \label{alphaEOM}
        \\[2ex]
        \nabla\times \vec{B} &=& -ie\left( \varphi^*\nabla\varphi - \varphi\nabla\varphi^* \right)
        -2e\left(e\bm{A} +\nabla\alpha\right)|\varphi|^2\,,
        \label{gaugeEOM}
    \eea
    \label{EOMs}%
\end{subequations}
where $\vec{B} = \nabla\times \vec{A}$ is the magnetic field and where we have defined the operator
\bea \label{Ddef}
{\cal D} &\equiv& \Delta - 2i(e\vec{A}+\nabla\alpha)\cdot\nabla-i\nabla\cdot(e\vec{A}+\nabla\alpha)-(e\vec{A}+\nabla\alpha)^2+(\nabla\alpha)^2-m_\pi^2\cos\alpha \, .\hspace{0.5cm}
\eea
Since we are ignoring the contributions of fluctuations to the thermodynamics, the grand canonical potential density is simply given by $\Omega=-{\cal L}$. Implementing our assumptions of a static system and vanishing electric field we obtain 
\bea
    \Omega(\vec{x}) &=& \frac{B^2}{2}+|\left[\nabla -i\left(e\bm{ A}+\nabla\alpha\right)\right]\varphi|^2 +\frac{\left(\nabla|\varphi|^2\right)^2}{2\left(f_{\pi}^2-2|\varphi|^2\right)} +\frac{f_{\pi}^2 -2|\varphi|^2}{2}\left(\nabla\alpha\right)^2
        \non[2ex]
        &&-m_{\pi}^2f_{\pi}\sqrt{f_{\pi}^2-2|\varphi|^2}\cos{\alpha}
        -\mu n_B(\vec{x})         \, , 
        \label{ThermoPotential}
\eea
where 
\begin{equation}
   n_{B}(\vec{x})= j_B^0 = \frac{\nabla\alpha}{4\pi^2}\cdot\left(e\vec{B}+\frac{\nabla\times\vec{j}}{ef_\pi^2}\right) 
   \label{nblocal}
\end{equation}
is the (local) baryon density. This result follows from the baryon current (\ref{BCurrent}) with the convention $\epsilon^{0123}=+1$, and the charged three-current is defined through $j^\mu = (j^0,\vec{j})$.   
With the help of the equation of motion (\ref{phiEOM}) we can write the 
free energy  as  
\bea
    F&=&\int d^3\vec{x}\, \Omega(\vec{x}) \,, \non[2ex]    
    &=&\int d^3\vec{x}\bigg[\frac{B^2}{2}+ \frac{f_{\pi}^2}{2}\left(\nabla\alpha\right)^2 -\frac{m_{\pi}^2 f_{\pi}\cos{\alpha}}{\sqrt{f_{\pi}^2-2|\varphi|^2}}\left(f_{\pi}^2 -|\varphi|^2\right)-\frac{f_{\pi}^2}{2}\frac{(\nabla|\varphi|^2)^2}{\left(f_{\pi}^2-2|\varphi|^2\right)^2} 
     -\frac{e\mu}{4\pi^2}\nabla \alpha \cdot \bm{B} \bigg]\non[2ex]
    && 
    +\int d\bm{S}\cdot\bigg\{\varphi^*\left[\nabla -i\left(e\bm{A}+\nabla\alpha\right) +\frac{\nabla|\varphi|^2}{f_{\pi}^2-2|\varphi|^2}\right]\varphi 
    +\frac{\mu(\nabla\alpha\times \bm{j})}{4\pi^2ef_{\pi}^2}   \bigg\}\,,
    \label{FreeEnergySimplified}
\eea
where we have separated the surface terms that we can drop in our evaluation later. We shall denote the resulting free energy density by 
\be
{\cal F} = \frac{F}{V} \, ,
\ee
where $V$ is the volume of the system.

%%%%%%%%%%%%%%%%%%%%%%%%%%%%%%%%%%%%%%%%%%%%%%%%%%%%%%%%%%%%%%%%%%%%%%%%%%%%%%%%%%%%%%%%%%%%%%%%%%%%%%%%%%%%%%%%%
\subsection{Instability at the critical magnetic field}
\label{sec:CSL}
%%%%%%%%%%%%%%%%%%%%%%%%%%%%%%%%%%%%%%%%%%%%%%%%%%%%%%%%%%%%%%%%%%%%%%%%%%%%%%%%%%%%%%%%%%%%%%%%%%%%%%%%%%%%%%%%%

Let us briefly recapitulate the solution of the equations of motion in the absence of charged pions and the instability of the resulting phase at a certain critical magnetic field. It will be instructive to compare this instability with the analogous calculation for an ordinary superconductor, which is laid out in Appendix \ref{appB}. In the absence of charged pions, $\varphi=0$, the potential (\ref{ThermoPotential}) reduces to
\be
    \Omega_{\varphi=0} = \frac{B^2}{2}+\frac{f_{\pi}^2}{2}\left(\nabla\alpha\right)^2 -m_{\pi}^2f_{\pi}^2\cos{\alpha} -\frac{e\mu}{4\pi^2}\nabla \alpha \cdot \bm{B}\, ,
    \label{CSLThermoPotential}
\ee
while the equation of motion for $\alpha$ (\ref{alphaEOM}) is
\be
   \Delta\alpha = m_{\pi}^2\sin{\alpha}\,.
\ee
In the chiral limit, the solution that minimises the potential is
\begin{equation}
    \nabla \alpha = \frac{e \mu \bm{B}}{4\pi^2 f_{\pi}^2} \,,
    \label{GradAlpha}
\end{equation}
with corresponding free energy density 
\begin{equation}
    {\cal F}_{\rm CSL} = \frac{\langle B\rangle^2}{2}-\frac{1}{2} \left(\frac{e \mu \langle B\rangle}{4\pi^2 f_{\pi}}\right)^2\,.
    \label{FreeEnergyAlpha}
\end{equation}
In this phase, the magnetic field is uniform (and trivially fulfils the equation of motion (\ref{gaugeEOM})). Nevertheless, we have replaced it by its spatial average, defined for any function $f(\vec{x})$ by 
\be \label{avdef}
\langle f \rangle \equiv \frac{1}{V}\int d^3\vec{x}\, f(\vec{x}) \, .
\ee
Here we simply have $\langle B\rangle=B$, but in the form (\ref{FreeEnergyAlpha}) the free energy density can be compared more easily with our main results, where $B$ is no longer uniform. The baryon density in the chiral limit is also uniform and is given by 
\be \label{nBCSL}
n_B^{\rm CSL} = -\frac{\partial \Omega_{\varphi=0}}{\partial \mu} = \frac{e^2\mu\langle B\rangle^2}{16\pi^4 f_\pi^2} \, .
\ee
If a nonzero pion mass is taken into account, this result gets more complicated. In particular, $\nabla \alpha$ and the resulting baryon number vary periodically in the direction parallel to the magnetic field. This is the phase that was termed CSL \cite{Brauner:2016pko}.  Since our main results will only concern the chiral limit, the results (\ref{GradAlpha}) and (\ref{FreeEnergyAlpha}) are sufficient for our purposes. We have used the label CSL for notational convenience, although this is a slight abuse of the term since there is no lattice structure in the chiral limit in the direction of the magnetic field. 

The stability of the CSL phase can be probed by considering the fluctuations in the field $\varphi$. To this end, we go back to the equation of motion (\ref{phiEOMFull}). Setting $A_0=\partial_t\alpha=0$ but keeping the time dependence of $\varphi$ and linearising this equation yields
\bea
     0&\simeq&
     \left(\partial_{t}^2-{\cal D}  -\frac{ie\nabla\alpha\cdot\bm{B}}{4\pi^2f_{\pi}^2}\,\partial_{t}\right)\varphi 
     \, .
     \label{LOInstability}
\eea
To proceed we align the $z$-axis with the magnetic field, $\vec{B}=B\hat{\vec{e}}_z$, such that we can choose $e\bm{A}+\nabla\alpha = eBx\hat{\bm{e}}_y$. Moreover, we employ the ansatz $\varphi(t,\vec{x})= e^{-i\omega t}e^{ik_y y}f(x,z)$ to obtain 
\bea
     0= \left[-\omega^2 -\partial_x^2 - \partial_z^2+e^2B^2\left(x-\frac{k_y}{eB}\right)^2     -(\nabla\alpha)^2 +m_{\pi}^2\cos{\alpha}-\frac{e\nabla\alpha\cdot\bm{B}}{4\pi^2f_{\pi}^2}\, \omega\right]f(x,z)
     \, .\hspace{0.5cm} \label{ompartial}
\eea
Returning to the chiral limit, we abbreviate $\nabla\alpha = c\,\hat{\vec{e}}_z$ with 
\be \label{cdef}
c \equiv \frac{e \mu B}{4\pi^2 f_{\pi}^2}  \, ,
\ee
and further simplify the ansatz by writing $f(x,z)=e^{ik_{z}z} \psi(x)$.
This yields
\be
\left[(\omega+\mu_*)^2 - k_z^2 -m_*^2\right]\psi(x) = \left[-\partial_x^2+e^2B^2\left(x-\frac{k_y}{eB}\right)^2\right]\psi(x) \, .
\ee
Written in this form, this equation is identical to the standard Ginzburg-Landau scenario from $\varphi^4$ theory, see Eq.\ (\ref{omegapsi}), having identified an effective chemical potential  and an effective mass by 
\be
\mu_* = \frac{c^2}{2\mu} \, , \qquad m_*^2 = \mu_*^2 - c^2   \, .
\ee
Therefore, following exactly the same arguments as in Appendix \ref{appB}, the dispersion relation of the $\varphi$ field in the (massless) CSL phase is
\be\label{wstar}
\omega = \sqrt{(2\ell+1)eB+m_*^2+k_z^2}-\mu_* \, , 
\ee
and we encounter an instability of the $\ell=k_z=0$ mode for $eB< \mu_*^2-m_*^2=c^2$.
However, crucially, $\mu_*$ and $m_*$ depend on the magnetic field themselves. As a consequence, this condition translates into an instability for magnetic fields {\it larger} than the critical field 
\begin{equation}
    B_{c2}= \frac{16\pi^4f_{\pi}^4}{e\mu^2}\,.
    \label{Bc2}
\end{equation}
This is in contrast to the scenario of an ordinary type-II superconductor where the instability towards a superconducting flux tube lattice occurs upon {\it decreasing} the magnetic field. 

The critical magnetic field  (\ref{Bc2}) reproduces  the result of Ref.\ \cite{Brauner:2016pko} (where $e$ was set to 1 and the 
field was termed $B_{\rm BEC}$, indicating Bose-Einstein condensation of charged pions). In this reference, the critical field was also computed for the case of a nonzero pion mass (we have used this result in the phase diagram of Fig.\ \ref{fig:eBmu}).  Our derivation deviates in one detail from that of Ref.\ \cite{Brauner:2016pko}: The anomalous contribution in Eq.\ (\ref{phiEOMFull}) generates the term linear in $\omega$ in Eq.\ (\ref{ompartial}), which we then have absorbed in the effective chemical potential $\mu_*$. It is possible to discard this term on the ground of a consistent power counting scheme. As argued in Ref.\ \cite{BraunerKolesovaYamamoto:2021WarmCSL}, in addition to the usual power counting in chiral perturbation theory in terms of the momentum scale $p\ll 4\pi f_\pi$, namely $\partial_\mu, m_\pi,A_\mu \sim {\cal O}(p)$, the baryon chemical potential should be counted as $A_\mu^B \sim {\cal O}(p^{-1})$. This ensures that the contribution $A_\mu^B j_B^\mu \sim {\cal O}(p^2)$
in the WZW Lagrangian (\ref{ActionWZW}) is consistent with our chiral Lagrangian ${\cal L}_{\Sigma} \sim {\cal O}(p^2)$. 
In contrast, the second WZW contribution $eA_\mu j_B^\mu \sim {\cal O}(p^4)$ is of higher order. This is the term that gives rise to the effective chemical potential $\mu_*$. If $\mu_*$ is set to zero we reproduce the dispersion relation of Ref.\ \cite{Brauner:2016pko} exactly. However, since we also include the electromagnetic contribution ${\cal L}_{\rm em}\sim {\cal O}(p^4)$, which is crucial for our main results, 
our expansion is not consistent with respect to this scheme even in the absence of the WZW term.  Therefore, we have included all terms from ${\cal L}_{\rm WZW}$ (\ref{ActionWZW}), resulting in a formally higher-order term in the equation of motion (\ref{phiEOMFull}). 
An alternative power counting with respect to the electromagnetic field, namely $A_\mu \sim {\cal O}(p^0)$, $e \sim {\cal O}(p)$ \cite{Gronli:2022cri}, ensures consistency of the electromagnetic and chiral parts of the Lagrangian. Then, all our terms {\it are} consistently of order $p^2$ if we omit $eA_\mu j_B^\mu \sim {\cal O}(p^4)$ in the WZW Lagrangian, which is of higher order also within this alternative scheme. 

For the location of the instability the term linear in $\omega$ has no consequence because the critical magnetic field is given by $\mu_*^2-m_*^2$, which is identical to $c^2$ irrespective of whether $\mu_*$ is set to zero or not. Interestingly, however, the nature of the instability is affected: from Eq.\ (\ref{wstar}) we see that $\omega$ turns negative at the critical magnetic field, whereas, if we set $\mu_*=0$ in that equation, $\omega$ turns {\it imaginary} at the critical magnetic field. 
These two cases are sometimes referred to as "energetic" and "dynamical" instabilities. Only a dynamical instability indicates a time scale on which the unstable modes grow, whereas an energetic instability can be turned dynamical if the system is allowed to exchange momentum with an external system, see for instance Refs.\ \cite{Haber:2015exa,Andersson:2019ezz}. We thus see that only in the presence of $\mu_*$, the nature of the instability is the same as in the Ginzburg-Landau treatment of a standard superconductor. In that case, as one can check with the help of Eq.\ (\ref{wstar}), $\omega$ can only become complex in a regime which is already energetically unstable. For the following, this aspect of the instability as well as the difference in the dispersion of the charged pions in the CSL phase is irrelevant. In other words, if we count powers of the momentum scale according to $A_\mu^B \sim {\cal O}(p^{-1}), A_\mu \sim {\cal O}(p^0), e \sim {\cal O}(p)$ all our main results in the subsequent sections follow consistently from an order $p^2$ Lagrangian.

%%%%%%%%%%%%%%%%%%%%%%%%%%%%%%%%%%%%%%%%%%%%%%%%%%%%%%%%%%%%%%%%%%%%%

%%%%%%%%%%%%%%%%%%%%%%%%%%%%%%%%%%%%%%%%%%%%%%%%%%%%%%%%%%%%%%%%%%%%%
\section{Flux tube lattice}
\label{sec:EoMs}
%%%%%%%%%%%%%%%%%%%%%%%%%%%%%%%%%%%%%%%%%%%%%%%%%%%%%%%%%%%%%%%%%%%%%

%%%%%%%%%%%%%%%%%%%%%%%%%%%%%%%%%%%%%%%%%%%%%%%%%%%%%%%%%%%%%%%%%%%%%
\subsection{Expansion at the critical magnetic field} 
\label{sec:exp0}
%%%%%%%%%%%%%%%%%%%%%%%%%%%%%%%%%%%%%%%%%%%%%%%%%%%%%%%%%%%%%%%%%%%%%

The instability discussed in the previous section indicates that there is a phase that includes charged pion condensation and  has lower free energy than the CSL phase for magnetic fields above the critical field $B_{c2}$. In this section, we construct such a phase by applying an expansion in the parameter $\epsilon \sim \sqrt{B-B_{c2}}$, exploiting the analogy with the standard type-II superconductor of Appendix \ref{appB}. We shall present the expansion for the general case, including the pion mass, but restrict ourselves to the chiral limit in the solution of the resulting equations in Sec.\ \ref{sec:chiral}. In contrast to ordinary gauged $\varphi^4$ theory, we have the additional scalar field $\alpha$, which we also have to expand,
\begin{equation} \label{expansion}
    \varphi  = \varphi_0 +\delta\varphi +\ldots \,, \qquad \alpha = \alpha_0 + \delta\alpha + \ldots \, , \qquad \bm{A} = \bm{A}_0 +\delta\bm{A} + \ldots  \, . 
\end{equation}
Here, $\varphi_0$ and $\delta\varphi$ are of order $\epsilon$ and $\epsilon^3$, respectively. To order $\epsilon^0$, the gauge field and the scalar field take the values of the CSL phase, here denoted by $\vec{A}_0$ and $\alpha_0$. 
The higher-order terms $\delta\bm{A}$ and $\delta\alpha$ are of order $\epsilon^2$. We insert these expansions into the equations of motion (\ref{EOMs}) to obtain the following order-by-order equations. From the equation of motion for $\varphi^*$ (\ref{phiEOM}) we obtain the 
$\epsilon^1$ and $\epsilon^3$ contributions
\begin{subequations}
\bea
{\cal D}_0\varphi_0&=&0  \,, \label{phi0EoMEx} \\[2ex]
{\cal D}_0\delta\varphi &=& \Bigg[2i(e\delta\vec{A}+\nabla\delta\alpha)\cdot\nabla+i\nabla\cdot(e\delta\vec{A}+\nabla\delta\alpha)+2(e\vec{A}_0+\nabla\alpha_0)\cdot(e\delta\vec{A}+\nabla\delta\alpha)\non[2ex]
&&-(\Delta\alpha_0+2\nabla\alpha_0\cdot\nabla)\delta\alpha-\frac{(\Delta-m_\pi^2\cos\alpha_0)|\varphi_0|^2}{f_\pi^2}\Bigg]\varphi_0 \,  ,\label{deltaphiEoMEx}
\eea
\end{subequations} 
where ${\cal D}_0$ is the lowest-order contribution to the operator ${\cal D}$, i.e. Eq.\ (\ref{Ddef}) with $\vec{A}$ and $\alpha$ replaced by $\vec{A}_0$ and $\alpha_0$. 
The equation of motion for $\alpha$ (\ref{alphaEOM}) yields the following $\epsilon^0$ and $\epsilon^2$ equations,
\begin{subequations}
\bea
\Delta\alpha_0 &=& m_\pi^2 \sin\alpha_0 \, , \label{alpha0EoMEx} \\[2ex]
f_\pi^2(\Delta-m_\pi^2\cos\alpha_0)\delta\alpha &=& (\Delta\alpha_0+2\nabla\alpha_0\cdot\nabla) |\varphi_0|^2 \, , \label{deltaalphaEoMEx}
\eea
\end{subequations}
where the first equation has already been used to simplify the second. 
Finally, from the equation of motion for $\vec{A}$, we derive the $\epsilon^0$ and $\epsilon^2$ contributions
\begin{subequations}
\bea
\nabla\times \vec{B}_0 &=& 0 \, , \label{A0EoMEx}\\[2ex]
\nabla\times\delta\vec{B} &=& -ie\left(\varphi_0^*\nabla\varphi_0 -\varphi_0\nabla\varphi_0^*\right) -2e\left(e\bm{A}_0 +\nabla\alpha_0\right)|\varphi_0|^2 \, , \label{deltaAEoMEx}
\eea
\end{subequations}
where we have defined $\vec{B}_0 = \nabla\times \vec{A}_0$, $\delta\vec{B} = \nabla\times\delta\vec{A}$, such that we can denote the magnetic field up to ${\cal O}(\epsilon^2)$ by $\vec{B} \simeq \vec{B}_0  + \delta\vec{B}$. The lowest-order magnetic field corresponds to the critical magnetic field, $B_0 = B_{c2}$ and we satisfy Eq.\ (\ref{A0EoMEx}) trivially by a constant $\vec{B}_0$. 

The free energy density  can be brought into a convenient form by dropping the surface terms and applying our expansion
in Eq.\ (\ref{FreeEnergySimplified}), 
\bea \label{FreeEnergyEx}
{\cal F} \simeq \frac{1}{V}\int d^3\vec{x}\left\{\frac{B^2}{2}+\frac{f_\pi^2}{2}[(\nabla\alpha)^2-2m_\pi^2\cos\alpha]-\lambda_*|\varphi_0|^4-\frac{e\mu}{4\pi^2}\nabla\alpha\cdot\vec{B}
\right\} \, ,
\eea
where we have defined 
\be\label{lamstar}
\lambda_*\equiv \frac{\langle 
(\nabla|\varphi_0|^2)^2\rangle +m_\pi^2\langle|\varphi_0|^4 \cos\alpha \rangle}{2f_\pi^2\langle|\varphi_0|^4\rangle} \, .
\ee
Written in this way, the free energy density resembles the one of $\varphi^4$ theory, see Eq.\ (\ref{appF}), which will be helpful for the upcoming evaluation. In particular, $\lambda_*$ plays the role of an effective self-coupling of the complex field.  

As in the $\varphi^4$ calculation of Appendix \ref{appB} we use the higher-order equations (\ref{deltaphiEoMEx}) and (\ref{deltaAEoMEx}) to derive an identity that will later be needed to evaluate the free energy. We multiply Eq.\ (\ref{deltaphiEoMEx}) by $\varphi^*$ and Eq.\ (\ref{deltaAEoMEx}) by $\delta\vec{A}+\nabla\delta\alpha/e$, and combine the resulting equations to obtain 
\bea
\varphi_0^*{\cal D}_0\delta\varphi &=& i\nabla\cdot\left[(e\delta\vec{A}+\nabla\delta\alpha)|\varphi_0|^2\right]-\left(\delta\vec{A}+\nabla\delta\alpha/e\right)\cdot(\nabla\times\delta\vec{B})\non[2ex]
&&-\frac{|\varphi_0|^2(\Delta-m_\pi^2\cos\alpha_0)|\varphi_0|^2}{f_\pi^2}-|\varphi_0|^2(\Delta\alpha_0+2\nabla\alpha_0\cdot\nabla)\delta\alpha\, .
\eea
Integrating over the volume on both sides and dropping the surface terms gives
\bea
0&=& \int d^3\vec{x}\left[2\lambda_*|\varphi_0|^4-(\delta\vec{A}+\nabla\delta\alpha/e)\cdot(\nabla\times\delta\vec{B}) -|\varphi_0|^2(\Delta\alpha_0+2\nabla\alpha_0\cdot\nabla)\delta\alpha\right] \, . \label{intAdB}
\eea
This is the analogue to Eq.\ (\ref{appBorthogonal}). The extra term due to the scalar field $\alpha$ will play an important role below.

%%%%%%%%%%%%%%%%%%%%%%%%%%%%%%%%%%%%%%%%%%%%%%%%%%%%%%%%%%%%%%%%%%%%%
\subsection{Solution in the chiral limit}
\label{sec:chiral}
%%%%%%%%%%%%%%%%%%%%%%%%%%%%%%%%%%%%%%%%%%%%%%%%%%%%%%%%%%%%%%%%%%%%%

We now solve the equations of motion and compute the free energy density in the chiral limit, $m_\pi=0$. According to our expansion, the lowest-order terms of $\alpha$ and $\vec{A}$ correspond to their CSL values at the critical field $B_{c2}$. Aligning the magnetic field with the $z$-direction, we can thus write 
\be\label{alpha0}
\alpha_0 = \frac{z}{\xi} \, , \qquad e\bm{A}_0+\nabla\alpha_0 = eB_{c2} x\hat{\bm{e}}_y \, , 
\ee
where $\xi^{-1}$ is the constant $c$ from Eq.\ (\ref{cdef}) evaluated at 
$B=B_{c2}$, 
\be\label{xidef}
\xi = \frac{1}{\sqrt{eB_{c2}}} = \frac{\mu}{4\pi^2f_\pi^2} \, , 
\ee
where we have used the explicit form of $B_{c2}$ (\ref{Bc2}). 
Our gauge choice for $\vec{A}_0$ implies $\vec{B}_0 = B_{c2} \hat{\vec{e}}_z$. Moreover, by assigning a $z$-component to $\vec{A}_0$ that absorbs $\nabla \alpha_0$ we ensure that there is no charged current in the $z$-direction, $j_z=0$, which is a convenient choice for the calculation. 

With Eq.\ (\ref{alpha0}) we can solve the 
equation of motion for $\varphi_0$ (\ref{phi0EoMEx}) in exactly the same way as for the standard superconductor, see Sec.\ \ref{app:sol}. The only difference is that $\mu^2-m^2$ from the $\varphi^4$ model is replaced by $1/\xi^2$. Consequently, following Appendix \ref{appB}, we have 
\be \label{phi0xy1}
    \varphi_0(x,y) =\sum_{n=-\infty}^{\infty}C_{n}e^{in qy}\psi_n(x)\,, \qquad \psi_n(x) = e^{-\frac{(x-x_n)^2}{2\xi^2}} \, , 
\ee
with complex coefficients $C_n$, the wave numbers $k_y=nq$, and $x_n \equiv nq\xi^2$. In particular, $\varphi_0$ does not depend on $z$ (which would be different in the presence of a pion mass because in that case $\nabla\alpha_0$ depends on $z$). The coherence length $\xi$, which characterises the variation of the condensate in the $x$-$y$ plane, is the same length scale as in $\alpha_0$ (\ref{alpha0}).

Next, we determine $\delta\vec{A}$ from Eq.\ (\ref{deltaAEoMEx}). 
Again, we can follow exactly the same arguments as in Sec.\ \ref{app:sol}. We can choose a gauge in which  $\delta\vec{A} = \delta A_y \hat{\vec{e}}_y$ such that $\delta \vec{B} = \delta B \hat{\vec{e}}_z$ with $\delta B = \partial_x \delta A_y$, and find
\be
    \delta A_y = \left(\langle B\rangle -B_{c2} +e\langle |\varphi_0|^2\rangle\right) x - e\int dx\,|\varphi_0|^2\,.
    \label{deltaA}
\ee
As a consequence, the magnetic field varies in the $x$-$y$ plane
and its $z$-component is 
\be
B =B_{c2} + \delta B =  \langle B\rangle + e(\langle|\varphi_0|^2\rangle-|\varphi_0|^2) \, ,
\ee
where $\langle B\rangle$ will act as our external thermodynamic variable.

In the chiral limit, the ${\cal O}(\epsilon^2)$ equation of motion  (\ref{deltaalphaEoMEx}) becomes $\Delta\delta\alpha=0$. We may assume $\delta\alpha$ to be independent of $x$ and $y$, and fix the integration constants such that the scalar field $\alpha$ up to ${\cal O}(\epsilon^2)$ is identical to its CSL value (\ref{GradAlpha}) at the magnetic field $\langle B\rangle$, 
\be
\alpha_0+\delta\alpha = \frac{e\mu\langle B\rangle}{4\pi^2f_\pi^2} z \, .
\ee
This condition is satisfied by  
\be
\delta\alpha = e(\langle B\rangle - B_{c2})\xi z \, .
\ee
The higher-order correction $\delta\varphi$ can now in principle be calculated from Eq.\ (\ref{deltaphiEoMEx}). However, we shall not need the explicit result. We already extracted information from that equation in the derivation of the relation (\ref{intAdB}). To make use of this relation we first compute
\be \label{dAdda}
(\delta\vec{A}+\nabla\delta\alpha/e)\cdot(\nabla\times\delta\vec{B}) = -e\left(\langle B\rangle - B_{c2} +e \langle|\varphi_0|^2\rangle\right) |\varphi_0|^2 +e^2 |\varphi_0|^4 +\mbox{total derivatives} \, ,
\ee
and 
\be \label{phiDelta}
|\varphi_0|^2(\Delta\alpha_0+2\nabla\alpha_0\cdot\nabla)\delta\alpha = 2e|\varphi_0|^2(\langle B\rangle - B_{c2}) \, . 
\ee
Interestingly, the expression (\ref{phiDelta}), when added to Eq.\ (\ref{dAdda}), effectively flips the sign of $\langle B\rangle-B_{c2}$, and Eq.\ (\ref{intAdB}) can be brought into the form
\begin{equation}
    e\langle|\varphi_0|^2\rangle = \frac{\langle B\rangle-B_{c2}}{\left(2\kappa^2-1\right)\beta +1}\,,
    \label{phi0BKbeta}
\end{equation}
with the effective Ginzburg-Landau parameter
\be \label{kappaeff}
\kappa \equiv \frac{\sqrt{\lambda_*}}{e} = \frac{1}{\sqrt{2}\,ef_\pi\xi} \, , 
\ee
where we have used the definition of $\lambda_*$ (\ref{lamstar}) and the identity
\be \label{phi22}
\langle (\nabla |\varphi_0|^2)^2\rangle = \frac{\langle|\varphi_0|^4\rangle}{\xi^2} \, , 
\ee
which we prove in Appendix \ref{appC}. Moreover, $\beta$ is the same parameter as introduced by Abrikosov in the standard Ginzburg-Landau scenario, see Eq.\ (\ref{appBbeta}).
The left-hand side of Eq.\ \eqref{phi0BKbeta} is obviously positive, and thus the right-hand side must be positive too. For the denominator, we find with Eqs.\ (\ref{xidef}), (\ref{kappaeff}), and using $f_\pi=92.4\, {\rm MeV}$, that $2\kappa^2-1>0$ for all $\mu \lesssim 12\, {\rm GeV}$ and thus for all relevant $\mu$. Therefore, our result is only valid for $\langle B\rangle - B_{c2}\ge 0$. This reflects the fact that our charged pion superconductor occurs for magnetic fields larger than the critical field. Hence the sign flip in front of $\langle B\rangle - B_{c2}$ due to Eqs.\ (\ref{dAdda}) and (\ref{phiDelta}) was crucial. In the absence of the scalar field $\alpha$, the contribution (\ref{phiDelta}) is absent and the numerator on the right-hand side of Eq.\ \eqref{phi0BKbeta} becomes $B_{c2}-\langle B\rangle $, which is positive in the standard scenario, see Eq.\ (\ref{appBphi0BKbeta}). 

We may also use Eq.\ (\ref{phi0BKbeta}) to determine the coefficients $C_n$ in the charged pion condensate (\ref{phi0xy1}).  As explained in  Sec.\ \ref{sec:lattice}, we consider periodic solutions where $C_n = C$ for even $n$ and $C_n=iC$ for odd $n$. Then, comparing Eq.\ (\ref{phisq1series}) with Eq.\ (\ref{phi0BKbeta}) we read off
\be \label{Csq}
|C|^2 = \frac{\sqrt{a}}{e}\frac{\langle B\rangle-B_{c2}}{\left(2\kappa^2-1\right)\beta +1} \, , 
\ee
where $a=q^2\xi^2/\pi$ determines the lattice structure of the solution. As a consistency check, this result shows that $|\varphi_0|^2$ is of order $\epsilon^2\sim \langle B\rangle-B_{c2}$, in accordance with our expansion (\ref{expansion}).  

Our results can now be inserted into the free energy density (\ref{FreeEnergyEx}). The magnetic energy and the $|\varphi_0|^4$ contribution have exactly the same form as in the $\varphi^4$ model, and we can use the result (\ref{appBFreeEnergyFinal}) for these terms. The remaining terms simply reproduce the free energy density of the CSL phase (\ref{FreeEnergyAlpha}), such that we obtain 
\be
    {\cal F} =  {\cal F}_{\rm CSL} - \frac{1}{2}\frac{\left(\langle B\rangle-B_{c2}\right)^2}{\left(2\kappa^2-1\right)\beta +1}\,.
    \label{FreeEnergyFinal}
\ee
This is one of the main results of our paper since it shows that the free energy of the inhomogeneous charged pion superconductor is indeed lower than that of the (massless) CSL state for $\langle B\rangle >B_{c2}$. Let us now discuss this result and the properties of our flux tube lattice in more detail. 

\begin{figure}
    \centering
   \hbox{\includegraphics[width=0.49\textwidth]{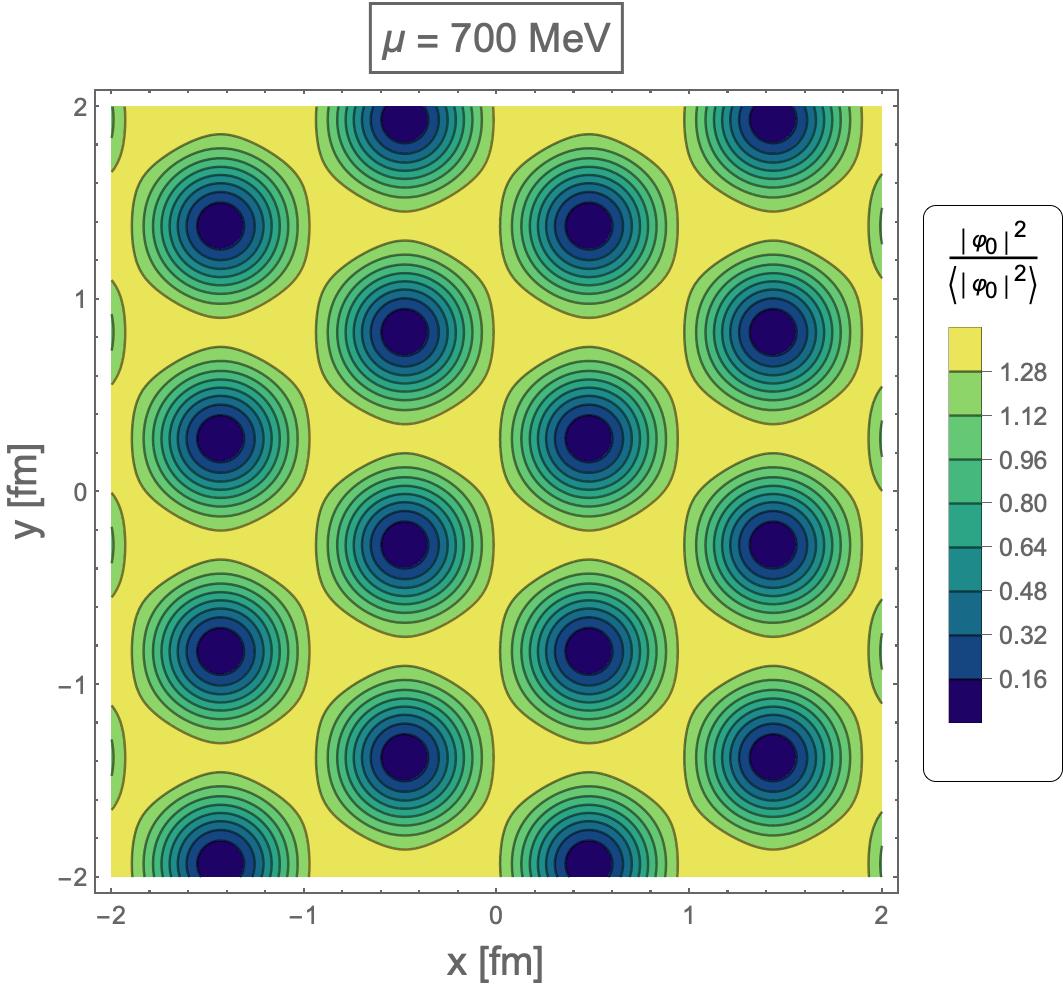} \hspace{0.2cm}\includegraphics[width=0.49\textwidth]{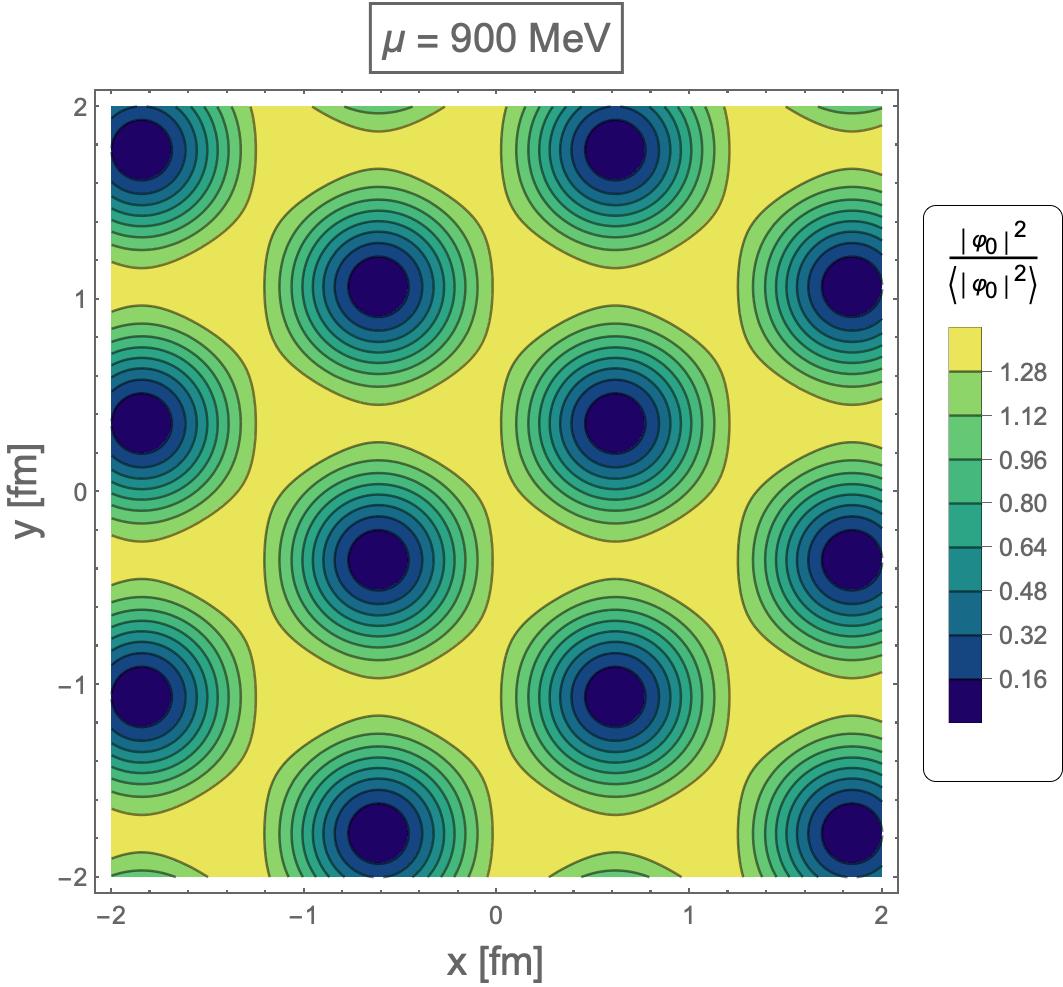}}
   \hbox{\includegraphics[width=0.49\textwidth]{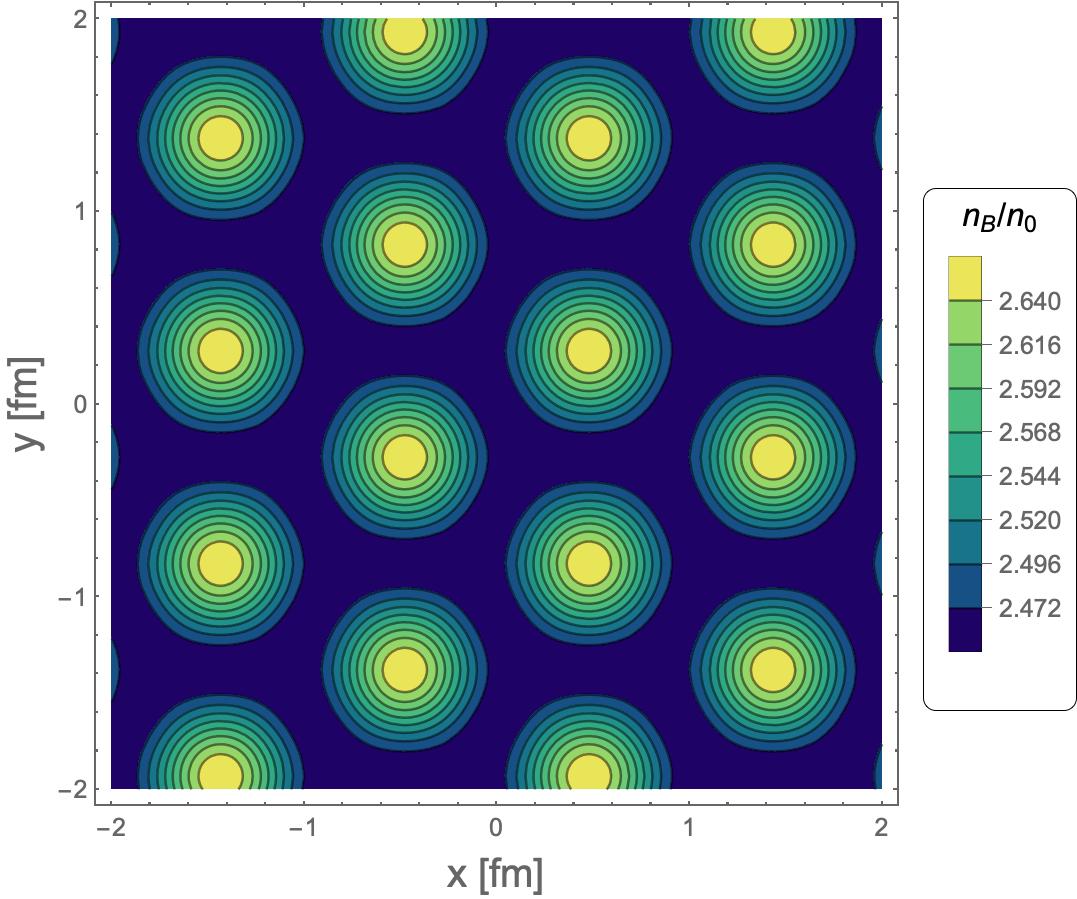} \hspace{0.2cm}\includegraphics[width=0.49\textwidth]{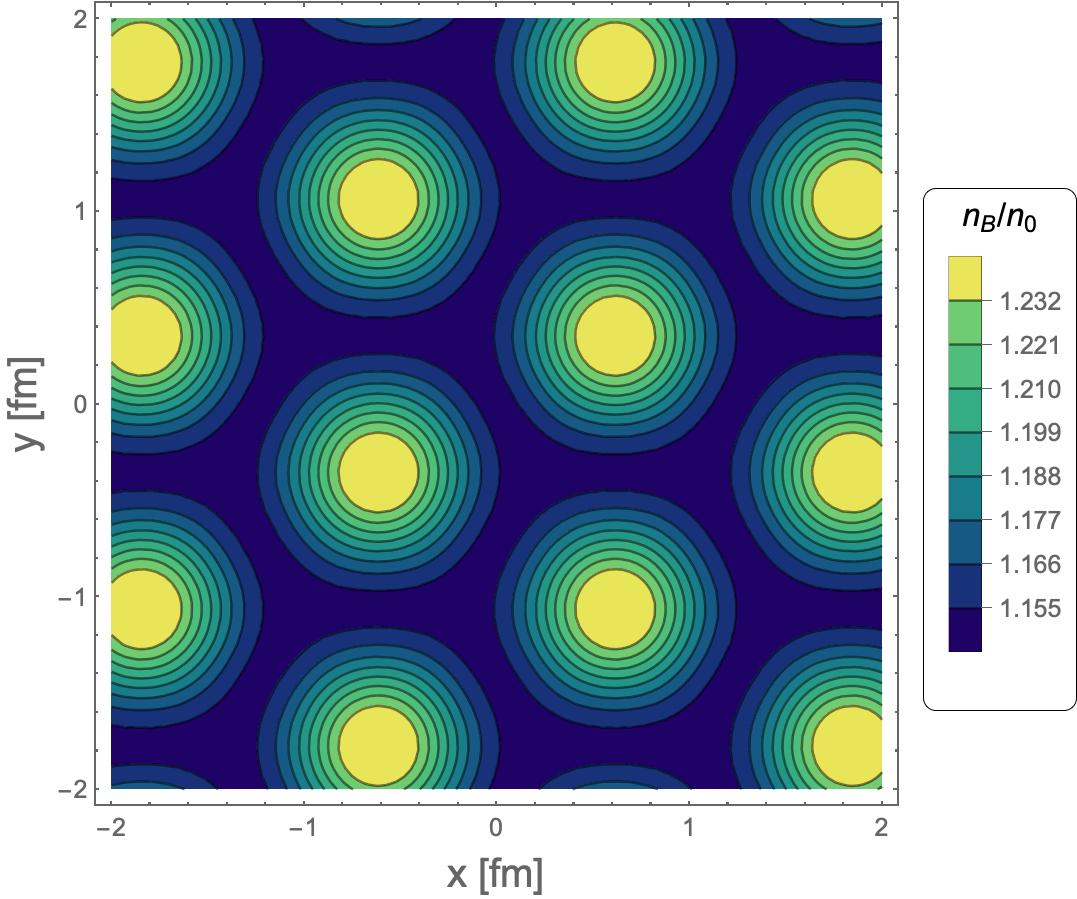}}
    \caption{{\it Upper panels:} Modulus of the charged pion condensate (squared) normalised to its spatial average in the plane perpendicular to the magnetic field.  The plots show the energetically preferred hexagonal lattice structure just above the critical magnetic field, in both panels $\langle B\rangle = 1.01 B_{c2}$, where $eB_{c2}\simeq 0.2319\, {\rm GeV}^2$ for a baryon chemical potential $\mu=700\, {\rm MeV}$ (left panel) and $eB_{c2}\simeq 0.1403\, {\rm GeV}^2$ for $\mu=900\, {\rm MeV}$ (right panel). The distance between the minima turns out to be about $1.10\, {\rm fm}$ (left) and $1.42\, {\rm fm}$ (right). {\it Lower panels:} Baryon number density in units of the nuclear saturation density $n_0$ for the same $\mu$ and $\langle B\rangle$ as the corresponding upper panels. We have used the numerical values $f_\pi=92.4\, {\rm MeV}$ and $n_0=0.15\, {\rm fm}^{-3}$. 
    }  \label{fig:BaryonNumberDensity}
\end{figure}

The inhomogeneous state we have constructed is preferred over the CSL state above the black curve in Fig.\ \ref{fig:eBmu}. This results in a continuous transition in the sense that the charged pion condensate goes to zero as the curve is approached from above. 
The charged pion condensate gives rise to a crystalline structure in the plane perpendicular to the magnetic field, while there is no variation of any physical observable in the direction parallel to it. The magnetic field itself varies in the $x$-$y$ plane as well, just like in an ordinary type-II superconductor. It is identical to the external field $\langle B\rangle$ at the points where the condensate vanishes and is expelled in the regions with nonvanishing condensate. The result is a flux tube lattice, whose structure is determined by the parameter $\beta$. Since the free energy is minimised by the minimal $\beta$ the preferred lattice structure is given by the minimum of the function (\ref{betaa}). There is no difference in this function to the case of an ordinary type-II superconductor, and thus we find the same result, i.e.\ the free energy is minimised by a hexagonal flux tube lattice, $a=\sqrt{3}$. Using the expression (\ref{phi0xy1}) and the coefficients (\ref{Csq}) we  plot the condensate for two different points in the $\mu$-$\langle B\rangle$ plane in the upper panels of Fig.\ \ref{fig:BaryonNumberDensity}. The amplitude of the oscillation becomes larger as one moves away from the critical field and our expansion in $\epsilon$ becomes less applicable. We have thus chosen two points very close to the critical field, where we can trust our expansion,  $(\mu,\langle B\rangle) = (\mu,1.01 \,B_{c2}(\mu))$, for two different values of $\mu$. 
These points are marked by diamonds in the phase diagram of Fig.\ \ref{fig:eBmu}.

We see that the lattice spacing increases with $\mu$. This is obvious since the characteristic length scale is the coherence length $\xi \propto \mu$. The lattice spacing also becomes larger as one moves away from the critical field $B_{c2}$. This is identical to an ordinary type-II superconductor, where for $\langle B\rangle \to 0$ (or $H\to H_{c1}$) the spacing becomes infinite, indicating a transition to a Meissner state where the magnetic field is completely expelled. Here, such a state is not possible. A homogeneous charged pion condensate would expel the magnetic field completely. However, without magnetic field there would be no anomalous coupling to the neutral pions and in turn there is no effective potential that makes the charged pions condense. Therefore, we do not expect our lattice to be continuously connected to a Meissner state, at least not in the absence of an isospin chemical potential. It is therefore not obvious -- even if we keep using chiral perturbation theory for such large magnetic fields and chemical potentials -- how the flux tube lattice evolves far beyond the critical field $B_{c2}$.   

Our crystalline state is not only a lattice for the charged pion condensate and thus the magnetic field, but also for baryon number due to the WZW term. The local baryon number (\ref{nblocal}) receives contributions both from $\nabla\alpha \cdot\vec{B}$ and $\nabla\alpha\cdot(\nabla\times \vec{j})$. With the help of Eqs.\ (\ref{iphi0}) and the definition of the (non-anomalous) charge current (\ref{QCurrent}) we have $\vec{j} \simeq -e(\hat{\vec{e}}_x\partial_y - \hat{\vec{e}}_y\partial_x)|\varphi_0|^2$, which implies 
\be
\nabla\times \vec{j} \simeq e\Delta|\varphi_0|^2 \hat{\vec{e}}_z \, .
\ee
Consequently, the baryon density (\ref{nblocal}) becomes
\be \label{nBxy}
n_B(x,y) \simeq n_B^{\rm CSL} + \frac{e^3\mu\langle B\rangle}{16\pi^4 f_\pi^2}\left[\langle|\varphi_0|^2\rangle-|\varphi_0|^2+\frac{\Delta|\varphi_0|^2}{e^2f_\pi^2}\right] \, , 
\ee
with the uniform CSL density $n_B^{\rm CSL}$ (\ref{nBCSL}), around which the baryon density oscillates, $\langle n_B \rangle = n_B^{\rm CSL}$. We plot the result in the lower panels of Fig.\ \ref{fig:BaryonNumberDensity} for the same points in the $\mu$-$\langle B\rangle$ plane as used in the upper two panels. We see that going above the critical field by 1\% leads to a periodic oscillation in the baryon number by about 8\%. The largest effect comes from the last term in Eq.\ (\ref{nBxy}), which originates from the vorticity contribution $\nabla\times \vec{j}$. This can be seen numerically or by estimating the ratio $[\Delta |\varphi_0|^2/(e^2 f_\pi^2)]/|\varphi_0|^2\sim 2\kappa^2 \gg 1$, where we have simply replaced the derivative by the inverse coherence length. We have checked numerically that this ratio is a good estimate for the relative importance of the two terms.

By comparing the upper with the lower panels, Fig.\ \ref{fig:BaryonNumberDensity} also demonstrates that the baryon density is enhanced where the charged pion condensate is depleted. As a consequence, we obtain a "baryon crystal" where baryon number is maximised at the triangular points of the lattice, just like the magnetic field. This two-dimensional lattice translates to "baryon tubes" in three dimensions since our system is translationally invariant in the direction of the magnetic field. This is due to our approximation of a vanishing pion mass. In the physical case, we can expect a three-dimensional crystal with baryon number oscillating in all three dimensions. As the phase diagram in Fig.\ \ref{fig:eBmu} suggests, the chiral limit becomes a good approximation for very large magnetic fields, $e\langle B\rangle \gg m_\pi^2$. Therefore, one might expect a change in the structure of the phase from tube-like at ultra-large magnetic fields to bubble-like at more moderate fields. Of course, this is under the assumption of the validity of our chiral approach, which we expect to break down at sufficiently large magnetic fields and/or baryon chemical potentials.

%%%%%%%%%%%%%%%%%%%%%%%%%%%%%%%%%%%%%%%%%%%%%%%%%%%%%%%%%%%%%%%%%%%%%
\section{Summary and outlook}
\label{sec:out}

We have applied chiral perturbation theory with a Wess-Zumino-Witten term to construct an inhomogeneous phase of superconducting charged pions coexisting with a neutral pion supercurrent. This phase is preferred over the chiral soliton lattice, where charged pions are absent, at sufficiently large magnetic fields and baryon chemical potentials. We have employed an expansion close to the critical magnetic field, making use of the methods developed for an ordinary type-II superconductor within a Ginzburg-Landau approach. Restricting ourselves to the chiral limit for our main results, 
we have derived an analytical expression for the crystalline phase and its free energy density. As in the case of an ordinary superconductor, it turns out that the preferred structure is a hexagonal flux tube lattice, where the magnetic field penetrates the superconductor in the regions of small condensate. Due to the chiral anomaly, this lattice is at the same time also a baryon crystal, with baryon number being enhanced within the flux tubes. We have pointed out that the main contribution to the oscillations in baryon number come from the vorticity of the charged pions which couples to the baryon chemical potential anomalously. Within our approximation of a vanishing pion mass, the baryon crystal is two-dimensional, only varying in the directions perpendicular to the magnetic field. 

Including a nonvanishing pion mass is the most obvious extension of our work in the future. We have included the pion mass in the necessary equations, including the expansion at the critical magnetic field. Their solution -- for which one may have to resort to numerical methods -- can be expected to provide a three-dimensional baryon crystal. As a first step, it might be useful to construct the crystal starting from a single domain wall rather than from the full chiral soliton lattice. Our results may also be used as the foundation to compute the baryon crystal purely numerically, without restriction to the region close to the critical magnetic field. In the presence of a finite pion mass, this would be particularly interesting in order to compute a potential first-order transition from the vacuum to the baryon crystal (without the intermediate state of the chiral soliton lattice), as suggested by the currently known structure of the phase diagram in the plane of magnetic field and baryon chemical potential. One should also keep in mind that our calculation resides near the limits  of validity of chiral perturbation theory, and thus any extensions beyond that approach would be highly desired to check and possibly refine our results. While a full first-principle calculation within QCD seems very difficult, one feasible extension would be the inclusion of actual baryonic matter. This would be relevant for the region of large chemical potential and relatively small magnetic fields, and one might expect a competition or a possible coexistence of  ordinary nuclear matter with our baryon crystal. The corresponding region of the phase diagram is also of potential interest for the interior of neutron stars, and one might ask if the crystalline structure discussed here might survive in some form in dense nuclear matter with moderately large magnetic fields, perhaps of the strengths found in magnetars. Moreover, one could include temperature effects along the lines of Ref.\ \cite{BraunerKolesovaYamamoto:2021WarmCSL}. Inhomogeneous pion condensates in a magnetic field can also be induced by an isospin chemical potential without anomalous effects \cite{Adhikari:2018Solo, Gronli:2022cri}. Therefore, it would be interesting to generalise our results by including an isospin chemical potential and see whether and how our crystalline structure connects to these known inhomogeneous phases.

\begin{acknowledgments}
We would like to thank Tom\'a\v{s} Brauner and Helena Kole\v{s}ov\'a for useful discussions and comments.
\end{acknowledgments}

%%%%%%%%%%%%%%%%%%%%%%%%%%%%%%%%%%%%%%%%%%%%%%%%%%%%%%%%%%%%%%%%%%%%%

\appendix

%%%%%%%%%%%%%%%%%%%%%%%%%%%%%%%%%%%%%%%%%%%%%%%%%%%%%%%%%%%%%%%%%%%%%
\section{Abrikosov flux tube lattice in $\varphi^4$ theory}
\label{appB}
%%%%%%%%%%%%%%%%%%%%%%%%%%%%%%%%%%%%%%%%%%%%%%%%%%%%%%%%%%%%%%%%%%%%%

In this appendix we discuss the second critical magnetic field and the resulting hexagonal flux tube lattice in a gauged $\varphi^4$ model. Despite the relativistic starting point, this essentially recapitulates the calculation of the original works \cite{Abrikosov:1957classic,Kleiner:1963Bulk}, in a notation adopted for our purposes. This calculation is useful as a warm-up for the more complicated version in the main part, and also serves to point out the crucial differences of our main results to the standard scenario. It also keeps the calculation in the main part to a more readable extent since we can resort to some of the results of this appendix. 

%%%%%%%%%%%%%%%%%%%%%%%%%%%%%%%%%%%%%%%%%%%%%%%%%%%%%%%%%%%%%%%%%%%%%
\subsection{Equations of motion and free energy}
%%%%%%%%%%%%%%%%%%%%%%%%%%%%%%%%%%%%%%%%%%%%%%%%%%%%%%%%%%%%%%%%%%%%%

We consider the following Lagrangian for a complex scalar field $\varphi$
with mass $m$, electric charge $e$, and coupling constant $\lambda$, 
\be
    \mathcal{L}=D_{\mu}\varphi \left(D^{\mu}\varphi \right)^* - m^2|\varphi|^2 -\lambda|\varphi|^4 - \frac{1}{4}F_{\mu \nu}F^{\mu \nu}\,,
    \label{ConventionalLagrangian}
\ee
where the covariant derivative is $D^{\mu}= \partial^{\mu} + ieA^{\mu}$. The equations of motion for $\varphi^*$ and $A^{\mu}$ are
\begin{subequations}
    \bea
        \left(D_{\mu}D^{\mu} +m^2 +2\lambda|\varphi|^2\right)\varphi&=&0\,,
        \label{appBphiEoM}
        \\[2ex]
        \partial_{\mu}F^{\mu\nu} &=&j^{\nu}\,,
    \eea
\end{subequations}
where 
\be
j^{\nu}= ie\left(\varphi^*\partial^{\nu}\varphi -\varphi\partial^{\nu}\varphi^*\right) -2e^2A^{\nu}|\varphi|^2
\ee
is the electromagnetic four-current. Condensation of the complex field is induced by a chemical potential $\mu$, which we introduce 
via the temporal component of the gauge field, $A^{\nu}=(\mu/e,\bm{A})$. Although we use the same symbol as for the baryon chemical potential in the main part, it is important to keep in mind the difference: in the case of the charged pions there is no chemical potential associated with the charge they carry. This would be an isospin chemical potential, which we do not consider in this paper. Their condensation only occurs through the coupling to the neutral pions, which in turn are coupled anomalously to the baryon chemical potential. Here, in this appendix, the condensation mechanism is more direct - $\mu$ is the chemical potential  associated to the global $U(1)$ symmetry of the model under which the complex field is charged.  

In the static limit the equations of motion become 
\begin{subequations}\label{appBEoM2}
    \bea
        0&=&\left({\cal D}  -2\lambda|\varphi|^2\right)\varphi\,, \label{appBEOM1}
        \\[2ex]
        \nabla\cdot\bm{E}  &=&  -2e^2\mu|\varphi|^2\,,
        \label{EOMDivE}
        \\[2ex]
        \nabla\times\bm{B} &=&  -ie\left(\varphi^*\nabla\varphi -\varphi\nabla\varphi^*\right) -2e^2\bm{A}|\varphi|^2 \, , 
    \eea 
\end{subequations}
where
\bea \label{calD}
{\cal D} &\equiv& \Delta-2ie\vec{A}\cdot\nabla-ie\nabla\cdot\vec{A}-e^2A^2+\mu^2-m^2 \, . 
\eea
Assuming that there is a background charge that cancels the charge of the complex scalar field, we will assume that there is no electric field, such that we can ignore Eq.\ (\ref{EOMDivE}) and do not have to take into account any electric contribution to the free energy. The free energy can then be written as 
\begin{equation}
    F=-\int d^3\vec{x}\, \mathcal{L} = \int d^3\vec{x} \left( \frac{B^2}{2} -\lambda|\varphi|^4\right)\,, \label{appF}
\end{equation}
where we have used the equation of motion (\ref{appBEOM1}) and dropped 
surface terms. 

%%%%%%%%%%%%%%%%%%%%%%%%%%%%%%%%%%%%%%%%%%%%%%%%%%%%%%%%%%%%%%%%%%%%%
\subsection{Expansion at the critical magnetic field}
%%%%%%%%%%%%%%%%%%%%%%%%%%%%%%%%%%%%%%%%%%%%%%%%%%%%%%%%%%%%%%%%%%%%%

At the critical field $B_{c2}$ we expect a continuous transition from the non-condensed phase $\varphi=0$ to a superconducting phase. To determine this transition, we linearise around $\varphi=0$ and temporarily restore the time dependence of $\varphi$. With the ansatz $\varphi(t,\vec{x})=e^{i\omega t}f(\vec{x})$ this allows us to compute the dispersion relation of the fluctuations in the non-superconducting state in the presence of the magnetic field. The equation of motion (\ref{appBphiEoM}) becomes 
\begin{equation}
    \left(\omega+\mu\right)^2f(\vec{x}) = -\left(\Delta -2ie\bm{A}\cdot\nabla -ie\nabla\cdot\bm{A}  -e^2A^2 -m^2\right)f(\vec{x})\,.
\end{equation}
Aligning the $z$-axis with the magnetic field, $\vec{B}=B\hat{\vec{e}}_z$, we may choose the gauge 
$\bm{A}=Bx\hat{\bm{e}}_y$ and make the ansatz $f(\vec{x})=e^{ik_{y}y}e^{ik_z z}\psi(x)$ to obtain 
\begin{equation}\label{omegapsi}
    \left[\left(\omega+\mu\right)^2-k_z^2-m^2\right]\psi(x) = \left[-\partial_{x}^2 +e^2B^2\left(x-\frac{k_y}{eB}\right)^2\right]\psi(x)\,.
\end{equation}
This equation has the form of the Schr\"{o}dinger equation for the one-dimensional harmonic oscillator and its solution gives the usual Landau levels labelled by the non-negative integer $\ell$, 
\begin{equation}
    \omega = \sqrt{\left(2\ell+1\right)eB +m^2 +k_{z}^2}-\mu\,.
\end{equation}
This energy  is positive for all $\ell$ and $k_z$ if $B$ is sufficiently large. A negative energy, and thus the indication of an instability, occurs for $\ell=k_z=0$ at the critical field
\begin{equation}
   B_{c2}  \equiv \frac{\mu^2-m^2}{e}\, . 
\end{equation}
For $B<B_{c2}$ we thus expect a superconducting phase with a charged condensate to take over. To construct this phase just below $B_{c2}$ we employ an expansion in $\epsilon\sim\sqrt{B_{c2}-B}$,
\begin{equation}
    \varphi  = \varphi_0 +\delta\varphi +\ldots \,, \qquad \bm{A} = \bm{A}_0 +\delta\bm{A} + \ldots \,,
\end{equation}
where, respectively, $\varphi_0$ and $\delta\varphi$ are of order $\epsilon$ and $\epsilon^3$, while $\vec{A}_0$ and $\delta\vec{A}$ are of order 1 and $\epsilon^2$. For $\epsilon\to 0$ we approach the critical field and thus we need $\nabla \times\vec{A}_0 = \vec{B}_{c2}$, which we can satisfy with $\vec{A}_0 = xB_{c2}\hat{\vec{e}}_y$. We may therefore write the expansion of the magnetic field as
\be
\vec{B} = \vec{B}_{c2} + \delta\vec{B} + \ldots \, , 
\ee
with $\nabla\times \delta\vec{A} = \delta\vec{B}$. The equation of motion for $\varphi^*$ (\ref{appBEOM1}) yields the order $\epsilon$ and $\epsilon^3$ equations
\begin{subequations}
\bea
{\cal D}_0 \varphi_0&=&0 \, , \label{appBphi0EoMEx} \\[2ex]
{\cal D}_0 \delta\varphi &=& \left(2ie\delta\vec{A}\cdot\nabla+2e^2\vec{A}_0\cdot\delta\vec{A}+ie\nabla\cdot\delta\vec{A}+2\lambda|\varphi_0|^2\right)\varphi_0 \, ,  \label{appBdeltaphiEoMEx}
\eea
\end{subequations}
where ${\cal D}_0$ is the operator ${\cal D}$ (\ref{calD}) with $\vec{A}$ replaced by its lowest order contribution $\vec{A}_0$. 
The lowest-order contribution to the equation of motion for $\vec{A}$ (\ref{appBEoM2}) simply gives $\nabla\times \vec{B}_{c2}=0$, which is trivially solved since the magnetic field is constant at (and above) the critical value. The order $\epsilon^2$ contribution gives
\be\label{appBdeltaAEoMEx}
\nabla\times\delta\vec{B} = -ie\left(\varphi_0^*\nabla\varphi_0 -\varphi_0\nabla\varphi_0^*\right) -2e^2\bm{A}_0|\varphi_0|^2 \, .
\ee
For later it is useful to combine Eq.\ (\ref{appBdeltaphiEoMEx}) with Eq.\ (\ref{appBdeltaAEoMEx}) as follows. We multiply Eq.\ (\ref{appBdeltaphiEoMEx}) from the left with $\varphi^*_0$ and multiply Eq.\ (\ref{appBdeltaAEoMEx}) 
with $\delta\vec{A}$. In both resulting equations we have created a term $2e^2\vec{A}_0\cdot\delta \vec{A}|\varphi_0|^2$, and thus we can insert one equation into the other to obtain
\be \label{phiDphi}
\varphi^*_0{\cal D}_0\delta\varphi = ie\nabla\cdot(\delta\vec{A}\,|\varphi_0|^2)-\delta\vec{A}\cdot(\nabla\times\delta\vec{B})+2\lambda|\varphi_0|^4 \, .
\ee
With partial integration,  dropping the surface term, and using the equation of motion ${\cal D}_0^*\varphi_0^*=0$, the integral over the left-hand side vanishes, 
\bea
\int d^3\vec{x} \, \varphi_0^*{\cal D}_0\delta\varphi &=& 0 \, .
\eea
Consequently, the integral over the right-hand side 
of Eq.\ (\ref{phiDphi}) must vanish as well. Dropping the boundary term, this yields the useful relation 
\bea \label{appBorthogonal}
0&=&\int d^3\vec{x}\left[2\lambda|\varphi_0|^4-\delta\vec{A}\cdot(\nabla\times\delta\vec{B})\right]  \, . 
\eea

%%%%%%%%%%%%%%%%%%%%%%%%%%%%%%%%%%%%%%%%%%%%%%%%%%%%%%%%%%%%%%%%%%%%%
\subsection{Solution to the equations of motion}
\label{app:sol}
%%%%%%%%%%%%%%%%%%%%%%%%%%%%%%%%%%%%%%%%%%%%%%%%%%%%%%%%%%%%%%%%%%%%%

To solve the equations of motion explicitly we first note that Eq.\ (\ref{appBphi0EoMEx}) can be brought into the form of Eq.\ (\ref{omegapsi}) with $\omega=0$,  $B=B_{c2}$, and, assuming no variation in the $z$-direction, $k_z = 0$. The solution for the lowest Landau level, $\ell=0$,  is a Gaussian, and in order to construct periodic solutions we set $k_y=n q$, $n\in \mathbb{Z}$, and consider the superposition of Gaussians
\be \label{phi0xy}
    \varphi_0(x,y) =\sum_{n=-\infty}^{\infty}C_{n}e^{in qy}\psi_n(x)\,, \qquad \psi_n(x) = e^{-\frac{(x-x_n)^2}{2\xi^2}} \, , 
\ee
with complex coefficients $C_n$ and the  abbreviations
\be
x_n \equiv n q\xi^2 \, , \qquad \xi\equiv \frac{1}{\sqrt{eB_{c2}}} \, .
\ee
Here, $\xi$ is the coherence length, which defines the length scale on which the condensate varies. 
Next, we need to compute $\delta\vec{A}$ and thus $\delta\vec{B}$ from the equation of motion (\ref{appBdeltaAEoMEx}). We will see that it is consistent to restrict the correction to the gauge field to the $y$ direction, $\delta \bm{A}=\delta A_y(x,y)\hat{\bm{e}}_y$, and we can write $\delta\vec{B} = \delta B(x,y)  \,\hat{\vec{e}}_z$. Now, one first derives the following useful identities with the help of the explicit solution (\ref{phi0xy}),
\begin{subequations} \label{iphi0}
\bea
i(\varphi_0^*\partial_x\varphi_0-\varphi_0\partial_x\varphi_0^*)&=& \partial_y|\varphi_0|^2 \, , \\[2ex]
i(\varphi_0^*\partial_y\varphi_0-\varphi_0\partial_y\varphi_0^*)+2e xB_{c2}|\varphi_0|^2 &=& -\partial_x|\varphi_0|^2 \, .
\eea
\end{subequations}
Consequently, the nontrivial components of Eq.\ (\ref{appBdeltaAEoMEx}) take the simple form
\begin{subequations}
\bea
\partial_y\partial_x \delta A_y&=& -e\partial_y|\varphi_0|^2 \, , \\[2ex]
\partial_x^2\delta A_y &=& -e\partial_x|\varphi_0|^2 \, .
\eea
\end{subequations}
The first equation gives $\partial_x\delta A_y = -e|\varphi_0|^2+{\rm const}$, and the second equation implies that the integration constant is indeed a constant that does not depend on $x$. We express the integration constant in terms of the spatial average of the magnetic field $\langle B\rangle$,  which we choose as our independent thermodynamic variable. Requiring $\langle B\rangle = B_{c2}+\langle \delta B \rangle$, we obtain
\be
\delta B = \partial_x\delta A_y = \langle B\rangle - B_{c2}+e\left(\langle|\varphi_0|^2\rangle-|\varphi_0|^2 \right) \, .
\ee
We can use this expression to compute 
\be
\delta\vec{A}\cdot(\nabla\times\delta\vec{B}) = -e\left(\langle B\rangle - B_{c2} +e \langle|\varphi_0|^2\rangle\right) |\varphi_0|^2 +e^2 |\varphi_0|^4 +\mbox{total derivatives} \, .
\ee
Inserting this result into Eq.\ (\ref{appBorthogonal}) and dropping the boundary terms gives
\be
    e\langle |\varphi_0|^2\rangle = \frac{B_{c2}-\langle B\rangle}{\left(2\kappa^2-1\right)\beta +1}\,,
    \label{appBphi0BKbeta}
\ee
where $\kappa\equiv \sqrt{\lambda}/e$ is the usual Ginzburg-Landau parameter  that distinguishes type-I from type-II superconductivity, and
\be
    \beta \equiv \frac{\langle|\varphi_0|^4\rangle}{\langle|\varphi_0|^2\rangle^2}\,.
    \label{appBbeta}
\ee
With these preparations we can now go back to the free energy (\ref{appF}). To express the result in terms of our thermodynamic variable $\langle B\rangle$, we need to rewrite the magnetic energy with the help of
\be
\langle B^2\rangle = \langle B\rangle^2+\langle\delta B^2\rangle-\langle\delta B\rangle^2 = \langle B\rangle^2 +e^2\left(\langle |\varphi_0|^4\rangle-\langle |\varphi_0|^2\rangle^2\right) \, ,
\ee
 such that, using Eq.\ (\ref{appBphi0BKbeta}), we obtain the free energy density  
\be
   {\cal F}\equiv \frac{F}{V}= \frac{\langle B\rangle^2}{2} -\frac{1}{2}\frac{\left(B_{c2} -\langle B\rangle\right)^2}{\left(2\kappa^2-1\right)\beta+1}\,.
    \label{appBFreeEnergyFinal}
\ee
This form is very useful since all the details of the lattice structure are captured by the parameter $\beta$, which was first introduced by Abrikosov \cite{Abrikosov:1957classic}. 

It is instructive to apply a Legendre transformation and instead of $\langle B\rangle$ use 
as a thermodynamic variable the external magnetic field
\be
H = \frac{\partial{\cal F}}{\partial\langle B\rangle} \, . 
\ee
With $H_{c2}=B_{c2}$ (since there is no magnetisation in the uncondensed state) this yields the Gibbs free energy
\be
{\cal G} = {\cal F} - \frac{1}{V}\int d^3\vec{x}\, H B  = -\frac{H^2}{2} -\frac{1}{2}\frac{(H_{c2}-H)^2}{(2\kappa^2-1)\beta} \, .
\ee
In this form we see that the free energy of the inhomogeneous condensed state is lower than that of the uncondensed state with Gibbs free energy $-H^2/2$ if and only if $\kappa>1/\sqrt{2}$ (since $\beta>0$), which is exactly the condition for type-II superconductivity. For $\kappa<1/\sqrt{2}$, in the type-I regime, the uncondensed phase undergoes a first-order phase transition at a critical field usually denoted by $H_c$ to a homogeneous superconductor where the magnetic field is expelled (i.e.\ $B=0$ although $H>0$). This discontinuous transition is not part of the present calculation because of the linearisation, which requires the condensate to be small.

%%%%%%%%%%%%%%%%%%%%%%%%%%%%%%%%%%%%%%%%%%%%%%%%%%%%%%%%%%%%%%%%%%%%%
\subsection{Lattice structures}
\label{sec:lattice}
%%%%%%%%%%%%%%%%%%%%%%%%%%%%%%%%%%%%%%%%%%%%%%%%%%%%%%%%%%%%%%%%%%%%%

 To compute the parameter $\beta$ for a given periodic structure in the $x$-$y$ plane we first introduce dimensionless variables with the help of the coherence length,
\be \label{dimless}
x\to \xi x  \, , \qquad y\to \xi y  \,, \qquad q\to \frac{q}{\xi} \, . 
\ee
Since the $z$ dependence of our system is trivial we can write the spatial average (\ref{avdef}) as 
\be
\langle f(x,y) \rangle =  \frac{1}{L_xL_y}\int_0^{L_x}dx\int_0^{L_y}dy\,f(x,y) \, .
\ee
We shall only be interested in configurations where  $L_x=2q$ and $L_y=2\pi/q$ \cite{Kleiner:1963Bulk}. We need to compute the spatial averages of $|\varphi_0|^2$ and $|\varphi_0|^4$ with $\varphi_0$ from Eq.\ (\ref{phi0xy}). In both cases, the $y$ integral produces a Kronecker delta such that we can write 
\begin{subequations}
\bea
\langle|\varphi_0|^2\rangle &=& \frac{1}{2q}\sum_n|C_n|^2\int_0^{2q}dx\,e^{-(x-qn)^2} \, , \label{phi21}\\[2ex]
\langle|\varphi_0|^4\rangle&=&\frac{1}{2q}\sum_{n,m,r}C_n^*C_mC_{m-n+r}^*C_r \int_0^{2q}dx\,e^{-2\left(x-\frac{m+r}{2}q\right)^2-\frac{q^2}{2}\left[(m-n)^2+(r-n)^2\right]} \non[2ex]
&=&\sum_{n,n_1,n_2}C_n^*C_{n+n_1}C_{n+n_1+n_2}^*C_{n+n_2}\, g^n_{n_1,n_2}\, , \label{phi41}
\eea
\end{subequations}
where each sum is over all integers from $-\infty$ to $\infty$. Also,  in Eq.\ (\ref{phi41}), we have introduced the new summation indices $n_1=m-n$ and $n_2=r-n$, and we have abbreviated 
\be
g^n_{n_1,n_2} \equiv \frac{e^{-\frac{q^2}{2}(n_1^2+n_2^2)}}{2q}  \int_0^{2q}dx\,e^{-2\left[x-\left(n+\frac{n_1+n_2}{2}\right)q\right]^2} \, . 
\ee
We shall now restrict ourselves to the periodic solutions $C_{n+2}=C_n$, such that we can write $C_n=C_0$ if $n$ is even and $C_n=C_1$ if $n$ is odd. One finds that all even terms in the infinite sum (\ref{phi21}) as well as all odd terms combine to give a Gaussian integral over $x\in[-\infty,\infty]$, such that 
\bea
\langle|\varphi_0|^2\rangle &=&\frac{\sqrt{\pi}(|C_0|^2+|C_1|^2)}{2q} \, . \label{phisq1series}
\eea
To evaluate Eq.\ (\ref{phi41}) we split the three summations into even and odd parts to write
\bea \label{CCCC}
\langle|\varphi_0|^4\rangle 
&=&\hspace{-0.2cm}\sum_{n,n_1,n_2}\hspace{-0.2cm}\Big[ |C_0|^4 g^{2n}_{2n_1,2n_2} + |C_1|^4g^{2n+1}_{2n_1,2n_2} +C_1^2(C_0^*)^2  g^{2n}_{2n_1+1,2n_2+1} + C_0^2(C_1^*)^2 g^{2n+1}_{2n_1+1,2n_2+1}  \non[2ex]
&&+ |C_0|^2|C_1|^2 \Big(g^{2n}_{2n_1,2n_2+1} + g^{2n}_{2n_1+1,2n_2} + g^{2n+1}_{2n_1,2n_2+1} +g^{2n+1}_{2n_1+1,2n_2}\Big)\Big] \, .
\eea
Now with $s=0,1$ and again piecing together the integration domains to obtain a Gaussian integral, we compute 
\bea
\sum_n g^{2n+s}_{n_1,n_2} &=& \frac{\sqrt{\pi}e^{-\frac{q^2}{2}(n_1^2+n_2^2)}}{2\sqrt{2}\,q} \, . 
\eea
Inserting this into Eq.\ (\ref{CCCC}) yields
\bea \label{phi4CC}
\langle|\varphi_0|^4\rangle &=& \frac{\sqrt{\pi}}{2\sqrt{2}\,q}\Big\{(|C_0|^4+|C_1|^4)f_0^2+4|C_0|^2|C_1|^2 f_0f_1+[C_1^2(C_0^*)^2+C_0^2(C_1^*)^2]f_1^2\Big\} \, , \hspace{1cm}
\eea
where
\be
f_s \equiv \sum_n  e^{-\frac{q^2}{2}(2n+s)^2} \, .
\ee
These functions can be expressed as  
\be
f_0 = \vartheta_3(0,e^{-2q^2}) \, , \qquad f_1 = e^{-q^2/2}\vartheta_3(iq^2,e^{-2q^2}) \, , 
\ee
with the Jacobi theta function 
\be \label{Jth}
\vartheta_3(z,x) = \sum_{n=-\infty}^\infty x^{n^2}e^{2niz}  \, .
\ee
Translational invariance $|\varphi_0(x,y)|^2 = |\varphi_0(x+(2N_x+N)L_x/2,y+(2N_y+N)L_y/2)|^2$ 
for any $N_x,N_y,N\in \mathbb{Z}$  is achieved by the choice 
\be
C_1=\pm iC_0 \label{C0C1} \, .
\ee
For this case, we have with the definition (\ref{appBbeta}) and the results (\ref{phisq1series}) and (\ref{phi4CC})
\bea \label{betaa}
\beta &=& \frac{q}{\sqrt{2\pi}}(f_0^2+2f_0f_1-f_1^2) \non[2ex]
&=& \sqrt{\frac{a}{2}} \Big\{ \left[\vartheta_3(0,e^{-2\pi a})\right]^2+2e^{-\frac{\pi a}{2}} \vartheta_3(0,e^{-2\pi a})\vartheta_3(i\pi a,e^{-2\pi a})\non[2ex]
&&-e^{-\pi a}\left[\vartheta_3(i\pi a,e^{-2\pi a})\right]^2\Big\}\, , 
\eea
where we have introduced the variable 
\be \label{aq}
a \equiv \tan\theta= \frac{L_x}{L_y} = \frac{q^2}{\pi} \, .
\ee
(Recall that $q$ is dimensionless here, in terms of dimensionful quantities $a=q^2\xi^2/\pi$.) This variable parametrises a continuum of triangular lattices, where $a=1$ and thus $\theta=\frac{\pi}{4}$ corresponds to a quadratic lattice, while $a=\sqrt{3}$ and thus $\theta = \frac{\pi}{3}$ gives a hexagonal lattice (or $a=1/\sqrt{3}$, which gives the same lattice with $x$ and $y$ directions swapped). As Eq.\ (\ref{appBFreeEnergyFinal}) shows, the preferred configuration is the one with the minimal $\beta$, and one finds that $\beta$ is minimised by the hexagonal structure, for which $\beta\simeq 1.1596$, while $\beta\simeq 1.1803$ for the quadratic lattice.

%%%%%%%%%%%%%%%%%%%%%%%%%%%%%%%%%%%%%%%%%%%%%%%%%%%%%%%%%%%%%%%%%%%%%
\section{Anomalous baryon current}
\label{appA}
%%%%%%%%%%%%%%%%%%%%%%%%%%%%%%%%%%%%%%%%%%%%%%%%%%%%%%%%%%%%%%%%%%%%%

In this appendix, we provide some details of the derivation leading to the result of the Goldstone-Wilczek baryon current in Eq.\ \eqref{BCurrent}. 
The starting point is the first line of Eq.\ (\ref{BCurrent}). Inserting the definition of the covariant derivative (\ref{covariant}) into this expression gives
\bea \label{jmuB1}
    j^{\mu}_{B} 
    &=& -\frac{\epsilon^{\mu\nu\rho\lambda}}{24\pi^2}\text{Tr}\left[-\Sigma\partial_{\nu}\Sigma^{\dagger}\partial_{\rho}\Sigma\partial_{\lambda}\Sigma^{\dagger} + \frac{3ie}{2}A_{\nu}\tau_3\left(\partial_{\rho}\Sigma^{\dagger}\partial_{\lambda}\Sigma -\partial_{\rho}\Sigma\partial_{\lambda}\Sigma^{\dagger} \right)\right.
    \non[2ex]
    &&\left.\hspace{1.9cm} +\frac{3ie}{4}F_{\nu\rho}\tau_3\left( \Sigma\partial_{\lambda}\Sigma^{\dagger} +\partial_{\lambda}\Sigma^{\dagger}\Sigma\right)\right]\,.
\eea
Next, we use the parametrisation given by Eqs.\ (\ref{SigmaU}) and (\ref{Usig}), $\Sigma = \Sigma_0U$ with $\Sigma_0=e^{i\alpha\tau_3}$.  One easily confirms $\partial_\nu\Sigma_0 = i\partial_\nu \alpha\, \tau_3 \Sigma_0$ and $[\Sigma_0,\tau_3]=[\Sigma_0^\dag,\tau_3]=0$. Moreover, one can check explicitly that $\tau_3U=U^\dag\tau_3$, and, since $U$ is unitary, we have $\partial_\mu U^\dag U =  - U^\dag\partial_\mu U$.
With the help of these relations the traces in Eq.\ (\ref{jmuB1}) become 
\begin{subequations}
\bea
\frac{\epsilon^{\mu\nu\rho\lambda}}{24\pi^2} \Tr[\Sigma\partial_\lambda\Sigma^\dag\partial_\nu\Sigma\partial_\rho\Sigma^\dag]&=&\frac{\epsilon^{\mu\nu\rho\lambda}}{24\pi^2}\Tr[U\partial_\lambda U^\dag\partial_\nu U\partial_\rho U^\dag]\non[2ex]
&& -\frac{i\epsilon^{\mu\nu\rho\lambda}}{8\pi^2}\partial_\nu\alpha\,\Tr[\tau_3 \partial_\rho U\partial_\lambda U^\dag] \, , \\[2ex]
-\frac{ie\,\epsilon^{\mu\nu\rho\lambda}}{16\pi^2} A_\lambda \Tr[\tau_3(\partial_\nu\Sigma^\dag\partial_\rho\Sigma-\partial_\nu\Sigma\partial_\rho\Sigma^\dag)]&=&-\frac{e\,\epsilon^{\mu\nu\rho\lambda}}{8\pi^2}A_\lambda \partial_\nu \alpha \,\Tr[U\partial_\rho U] \, , \\[2ex]
-\frac{ie\,\epsilon^{\mu\nu\rho\lambda}}{32\pi^2}F_{\nu\rho} \Tr[\tau_3(\Sigma\partial_\lambda\Sigma^\dag+\partial_\lambda\Sigma^\dag\Sigma)] &=&-\frac{e\,\epsilon^{\mu\nu\rho\lambda}}{32\pi^2}F_{\rho\lambda}\partial_\nu\alpha\,\Tr[1+U^2] \, .
\eea
\end{subequations}
These terms can be combined to the compact result
\be 
    j_B^\mu=\frac{\epsilon^{\mu\nu\rho\lambda}}{24\pi^2}\Tr[U\partial_\lambda U^\dag\partial_\nu U\partial_\rho U^\dag] +\partial_\nu G^{\mu\nu} \,, 
\label{appAjBtop}
\ee
where
\be
    G^{\mu\nu} = -\frac{\alpha \, \epsilon^{\mu\nu\rho\lambda}}{32\pi^2}\left(4i\Tr[\tau_3\partial_\rho U\partial_\lambda U^\dag]+4eA_\lambda\Tr[U\partial_\rho U]+e F_{\rho\lambda}\Tr[1+U^2]\right) \, .
    \label{appAGmunu}
\ee
Finally, by evaluating the traces we arrive at
\be
     j_{B}^{\mu}=-\frac{\epsilon^{\mu\nu\rho\lambda}}{4\pi^2}\partial_{\nu}\alpha\left\{\frac{e}{2}F_{\rho\lambda}+\frac{1}{f_{\pi}^2}\partial_{\rho}\left[ i\left(\varphi^*\partial_{\lambda}\varphi -\varphi\partial_{\lambda}\varphi^* \right) -2eA_{\lambda}|\varphi|^2\right] \right\}\,.
\ee
(The complex scalar field $\varphi$ is the rotated field of Eq.\ (\ref{phiprime}), but, as in the main part, we have dropped the prime for notational convenience.) One can now replace $eA_{\lambda}\rightarrow eA_{\lambda}-\partial_{\lambda}\alpha$ without changing the result and thus we arrive at the second line of Eq.\ \eqref{BCurrent}.

%%%%%%%%%%%%%%%%%%%%%%%%%%%%%%%%%%%%%%%%%%%%%%%%%%%%%%%%%%%%%%%%%%%%%
\section{Computing $\langle (\nabla |\varphi_0|^2)^2\rangle$}
\label{appC}
%%%%%%%%%%%%%%%%%%%%%%%%%%%%%%%%%%%%%%%%%%%%%%%%%%%%%%%%%%%%%%%%%%%%%

In this appendix we prove the identity (\ref{phi22}), which is needed for the effective coupling $\lambda_*$ in the calculation of the pion superconductor. We work with the dimensionless quantities (\ref{dimless}), such that with the form of the condensate (\ref{phi0xy1}) we find   
\bea
(\nabla |\varphi_0|^2)^2 &=& (\partial_x|\varphi_0|^2)^2+(\partial_y|\varphi_0|^2)^2 \non[2ex]
&=& \sum_{n,m,s,r} C_n^*C_mC_s^*C_r e^{i(m-n+r-s)qy} \psi_n(x)\psi_m(x)\psi_s(x)\psi_r(x)\non[2ex]
&&\times \left\{[2x-(m+n)q][2x-(s+r)q]-q^2(m-n)(r-s)\right\} \, .
\eea
As in Appendix \ref{sec:lattice} we work with periodic solutions and consider a rectangle in $x$ and $y$ with $L_x=2q$ and $L_y=2\pi/q$. Then, the spatial average becomes 
\bea
\langle(\nabla|\varphi_0|^2)^2\rangle &=&
\frac{1}{2q}\sum_{n,m,r}C_n^*C_mC_{m-n+r}^*C_r\int_0^{2q}dx\,e^{-2\left(x-\frac{m+r}{2}q\right)^2-\frac{q^2}{2}\left[(m-n)^2+(r-n)^2\right]}
\non[2ex]
&&\times \left\{4\left(x-\frac{m+r}{2}q\right)^2+q^2[m^2-r^2-2n(m-r)]\right\}\non[2ex]
&=&
\sum_{n,n_1,n_2}C_n^*C_{n+n_1}C_{n+n_1+n_2}^*C_{n+n_2}h^n_{n_1,n_2} \, , 
\eea
where we have employed the same renaming of summation variables as in Eq.\ (\ref{phi41}), and where we have abbreviated
\bea
h^{n}_{n_1,n_2} &\equiv& \frac{e^{-\frac{q^2}{2}(n_1^2+n_2^2)}}{2q}  \int_0^{2q}dx\,e^{-2\left[x-\left(n+\frac{n_1+n_2}{2}\right)q\right]^2}\non[2ex]
&&\times \left\{4\left[x-\left(n+\frac{n_1+n_2}{2}\right)q\right]^2+q^2(n_1^2-n_2^2)\right\} \, .
\eea
Assuming the same structure of the coefficients as in Appendix \ref{sec:lattice}, i.e.\ $C_n=C_{0/1}$ for $n$ even/odd, we obtain a sum analogous to Eq.\ (\ref{CCCC}),
\bea \label{CCCC1}
&&\langle(\nabla|\varphi_0|^2)^2\rangle \non[2ex]
&&=\hspace{-0.2cm}\sum_{n,n_1,n_2}\hspace{-0.2cm}\Big[ |C_0|^4 h^{2n}_{2n_1,2n_2} + |C_1|^4h^{2n+1}_{2n_1,2n_2} +C_1^2(C_0^*)^2  h^{2n}_{2n_1+1,2n_2+1} + C_0^2(C_1^*)^2 h^{2n+1}_{2n_1+1,2n_2+1}  \non[2ex]
&&+ |C_0|^2|C_1|^2 \Big(h^{2n}_{2n_1,2n_2+1} + h^{2n}_{2n_1+1,2n_2} + h^{2n+1}_{2n_1,2n_2+1} +h^{2n+1}_{2n_1+1,2n_2}\Big)\Big] \, .
\eea
By piecing together the infinite sum to Gaussian integrals we compute for $s=0,1$ 
\bea
\sum_n h^{2n+s}_{n_1,n_2} 
&=& \frac{\sqrt{\pi}e^{-\frac{q^2}{2}(n_1^2+n_2^2)}}{2\sqrt{2}q}\left[1+q^2(n_1^2-n_2^2)\right] \, .
\eea
Inserting this into Eq.\ (\ref{CCCC1}) we find that all contributions from the 
second term proportional to $n_1^2-n_2^2$ cancel each other. This can be seen by renaming the summation variables $n_1\leftrightarrow n_2$ suitably. Therefore, we find exactly the  same result as in Appendix \ref{sec:lattice}, see Eq.\ (\ref{phi4CC}), i.e.
\be \label{dPhiPhi}
\langle(\nabla|\varphi_0|^2)^2\rangle = \langle|\varphi_0|^4\rangle \, .
\ee
Since in the notation of this appendix the gradient denotes derivatives with respect to the dimensionless coordinates, we obtain Eq.\ (\ref{phi22}) after reinstating the coherence length $\xi$.

\bibliographystyle{JHEP}
\bibliography{references2}

\end{document}